\documentclass[journal, twocolumn]{IEEEtran}

\usepackage{cite}
\usepackage{amsmath,amssymb,amsfonts}
\usepackage{algorithmic}
\usepackage{graphicx}
\usepackage{textcomp}
\usepackage{subfigure}
\usepackage{booktabs}
\def\BibTeX{{\rm B\kern-.05em{\sc i\kern-.025em b}\kern-.08em
    T\kern-.1667em\lower.7ex\hbox{E}\kern-.125emX}}
\usepackage{cuted}
\usepackage{multirow}
\begin{document}

	\thispagestyle{empty}
	%\linenumbers  
	\clearpage

\title{Transmission Efficiency Limit of Single-Switch and Cascaded Reconfigurable Transmitarray Elements}
\author{Changhao Liu, \IEEEmembership{Graduate Student Member, IEEE}, Fan Yang, \IEEEmembership{Fellow, IEEE}, Shenheng Xu, \IEEEmembership{Member, IEEE} and Maokun Li, \IEEEmembership{Senior Member, IEEE}
	\thanks{This work was supported by \textit{(Corresponding author: Fan Yang.)}}
	\thanks{The authors are with the Department of Electronic Engineering, Tsinghua University, Beijing National Research Center for Information Science and Technology (BNRist),  Beijing 100084, China (e-mail: fan\_yang@tsinghua.edu.cn).}
}

\maketitle

\setcounter{page}{1}

\begin{abstract}
	Reconfigurable transmitarray antennas (RTAs) are rapidly gaining popularity, but optimizing their performance requires systematic design theories. In particular, establishing a performance limit theory for RTA elements is valuable. This paper presents a transmission efficiency limit theory for single-switch RTA elements and their cascaded extensions. Employing microwave network analysis, we analytically investigate single-switch RTA elements, demonstrating that their transmission coefficients under two states must lie on or within a specific unit circle on the Smith chart. Therefore, the transmission phase difference is tightly constrained by the transmission amplitudes, indicating that the phase-shifting ability of a single-switch RTA element is limited. Subsequently, this analysis is extended to cascaded RTA elements. By cascading several single-switch layers, the phase variation range is extended, enabling the realization of 1-bit phase shifts with high transmission amplitudes. These findings have significant impact on the design and optimization of RTA elements.
\end{abstract}


\section{Introduction}
\IEEEPARstart{R}{ecent} years have witnessed the rapid development of microwave metasurfaces, leading to the development of diverse high-gain array antennas with novel architectures, including reflectarray antennas (RAs), transmitarray antennas (TAs), and their reconfigurable versions \cite{reflectarray, RRA1, RRA2, RRA3}. Along with the emergence of various prototypes, underlying theories have also been proposed to guide the design of these array antennas. Performance limit theories are particularly useful in predicting the performance of general structures before their detailed design, which can significantly improve the design efficiency. Here are some examples of performance limit theories for these novel array antennas.

\begin{figure}[!t]
	\centering
	\subfigure[] {
		\includegraphics[width=0.28\columnwidth]{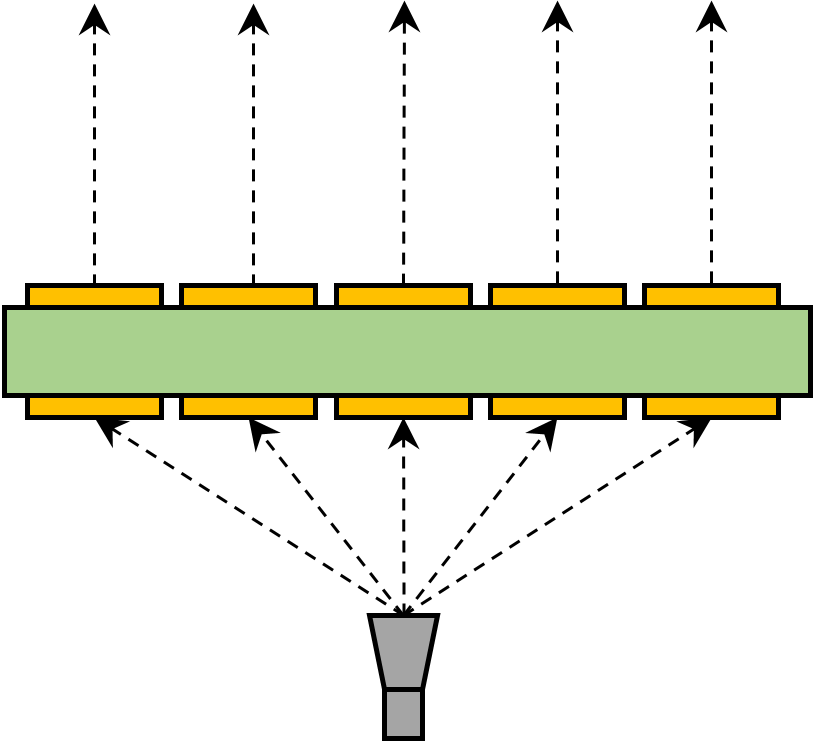} 
	} 
	\subfigure[] {
		\includegraphics[width=0.28\columnwidth]{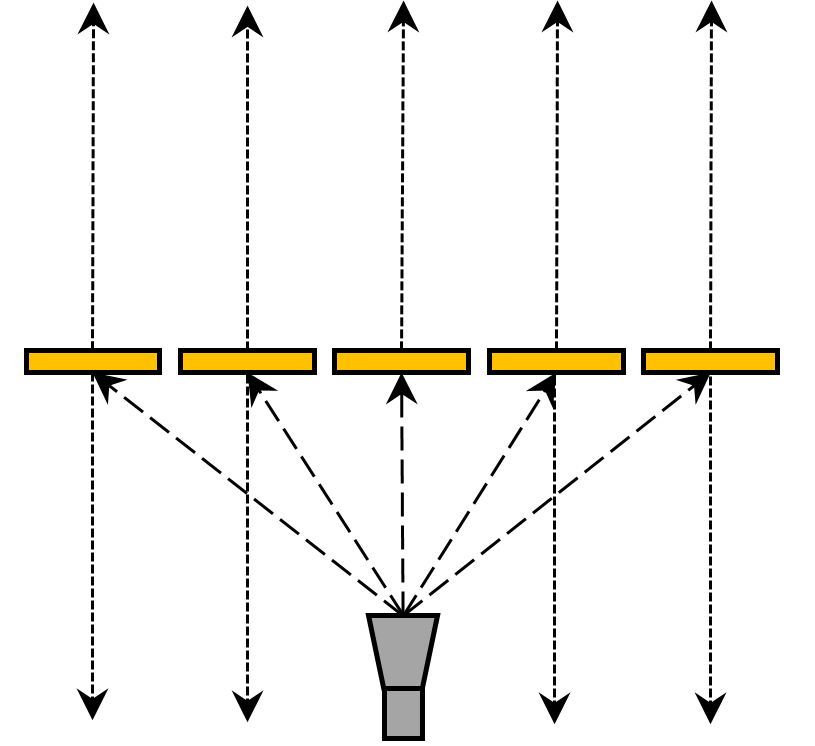} 
	} 
	\subfigure[] {
		\includegraphics[width=0.28\columnwidth]{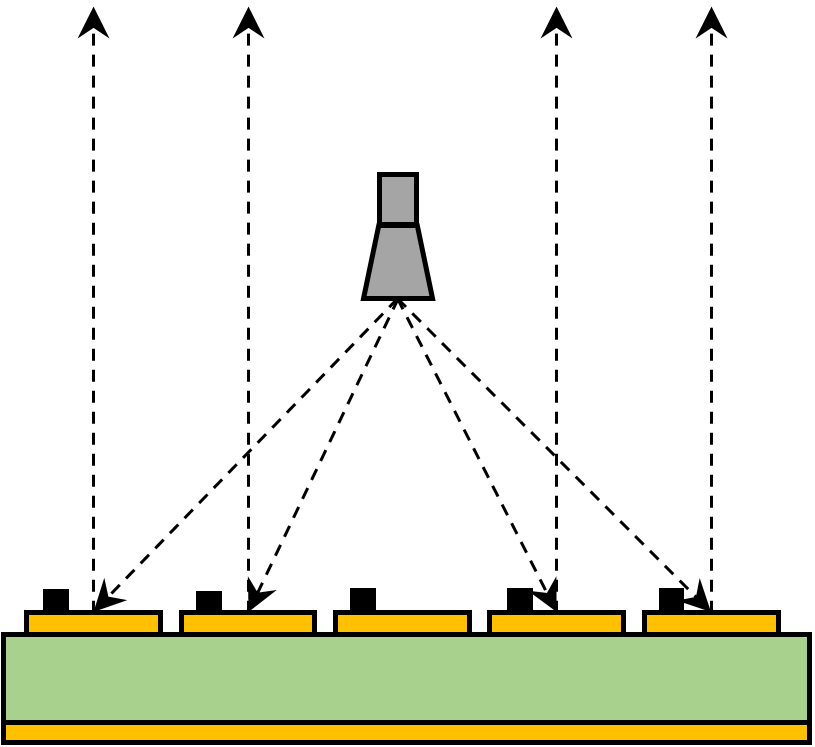} 
	} 
	\subfigure[] {
		\includegraphics[width=0.35\columnwidth]{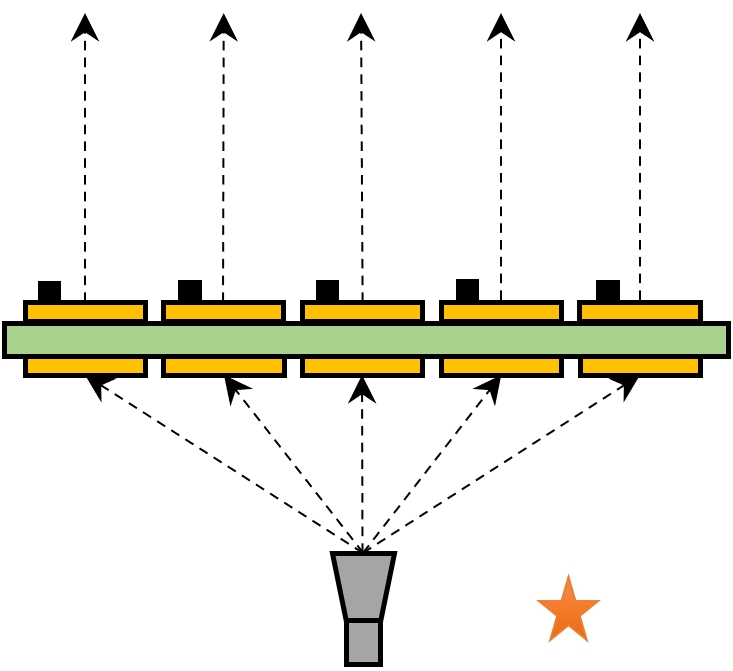} 
	} 
	\subfigure[] {
		\includegraphics[width=0.35\columnwidth]{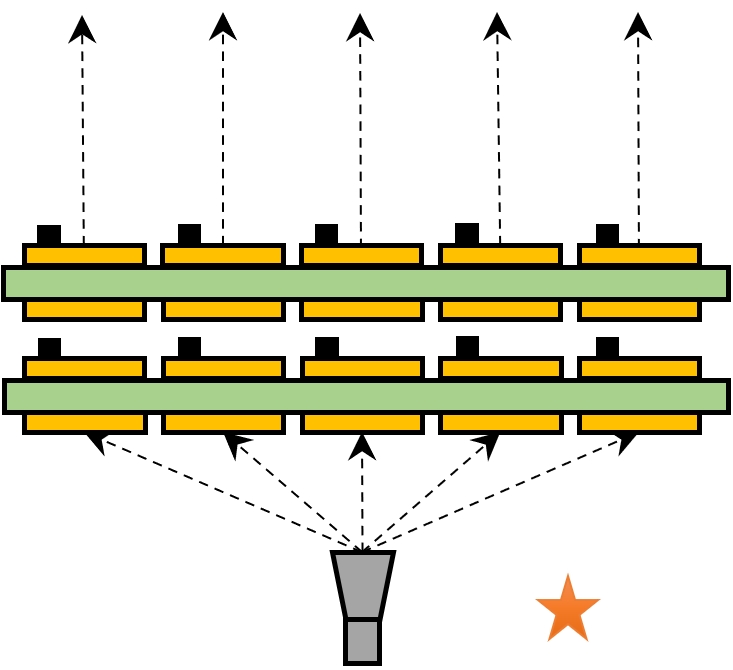} 
	} 
	\caption{Schematic illustrations of several novel array antennas. (a) Multi-layer TA. (b) Near-zero-thickness transmit-reflect-array antenna. (c) Single-switch RRA. (d) Single-switch RTA. (e) Cascaded double-layer RTA with one switch on each layer.}
	\label{arch}
\end{figure}

In 2014, the transmission phase limit theory for multilayer TA designs was introduced \cite{FSS}. Fig. \ref{arch}(a) depicts a schematic illustration of the multilayer TA. Using microwave network models and S-parameters, researchers discovered the constraints on the transmission amplitude and phase of TA elements, demonstrating that the phase tuning range within a 1-dB transmission amplitude loss is 54$^\circ$, 170$^\circ$, 308$^\circ$, and 360$^\circ$ for single-, double-, triple-, and quad-layers.

Another instance of the performance limit theory is transmit-reflect-array antennas based on polarization conversion methods, as revealed in 2018 \cite{tratheory}. A schematic of the this type of array antenna is shown in Fig. \ref{arch}(b). By applying the electric field continuity condition on the near-zero-thickness surface, researchers found that to achieve 360$^\circ$ phase coverage on both sides, a minimum of 6 dB amplitude loss is required.

Reconfigurable reflectarray antennas (RRAs) have emerged as a promising candidate for achieving beam-scanning capability. In particular, the 1-bit phase quantization technique has become a mainstream architecture in many applications, due to its minimal switch requirement with only 3 dB gain loss \cite{bit2}. In recent years, several single-switch 1-bit RRAs with different features have been presented in \cite{rra2,rra1,rra10,rra11,rra3,rra4,rra5,rra6,rra7,rra8,rra9,rra12}, as depicted in Fig. \ref{arch}(c). Recently, a performance limit theory for single-switch RRA elements was presented based on the radiation viewpoint \cite{rratheory,rratheory1}. By pre-calculating the performance limit and design targets at given switch parameters, this theory can facilitate the efficient design of RRA elements.

Reconfigurable transmitarray antennas (RTAs) have been proposed as an alternative to RRAs to eliminate the feed blockage effect. Since 2012, 1-bit RTAs have been extensively studied, and various prototypes based on different principles have been proposed. For example, current reversal method utilizes at least two opposite switches on one layer to achieve accurate 180$^\circ$ phase difference \cite{rta1,rta2,rta3,rta4,rta5,rta6,rta7,rta8,rta9}. Recently, stacked-layer RTAs with two switches on different layers have been presented based on the resonant method, which can also achieve 1-bit phase difference \cite{rta10,rta11,rta12}.

Interestingly, it has been found that all of the aforementioned 1-bit RTA elements require at least two switches to achieve 1-bit phase difference with high transmission amplitude. This raises a question: Is it feasible to design a high-performance RTA element using only one switch? Fig. \ref{arch}(d) provides an illustration of a single-switch RTA. Recently, by using three-port microwave network analysis, it is proven by contradiction that it is impossible to design a single-switch RTA element with both high transmission amplitudes and 1-bit phase tuning ability simultaneously \cite{rta, rtatheory}. Thus, theoretically, 1-bit RTA elements require a minimum of two switches. However, this proof only applies to the zero-reflection case, and it does not consider the general situation where the reflection coefficients are not necessarily zero. Therefore, a quantitative analysis of the performance of the single-switch RTA element in general cases needs to be further investigated.

\begin{figure}[!t]
	\centering
	\subfigure[] {
		\includegraphics[width=0.45\columnwidth]{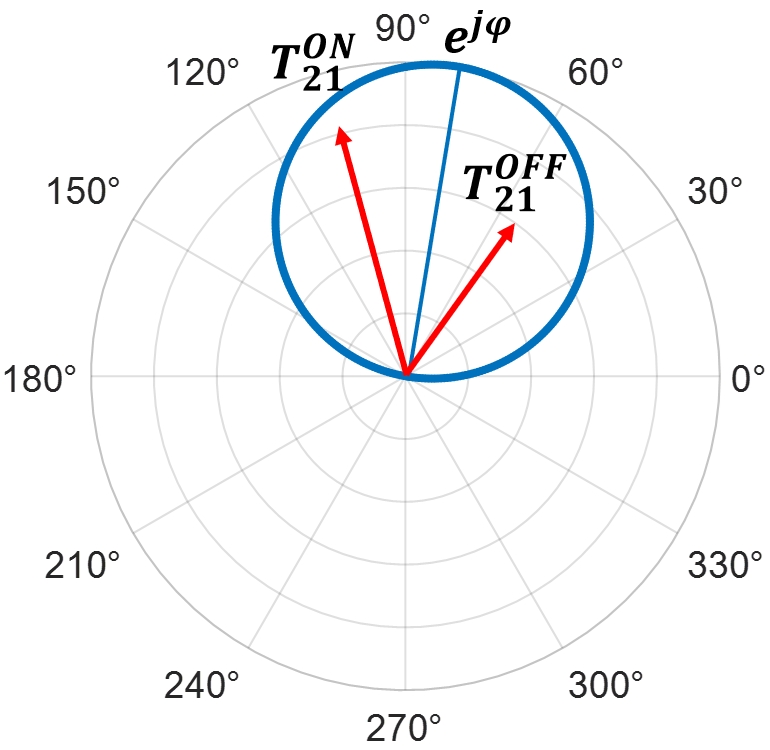} 
	} 
	\subfigure[] {
		\includegraphics[width=0.45\columnwidth]{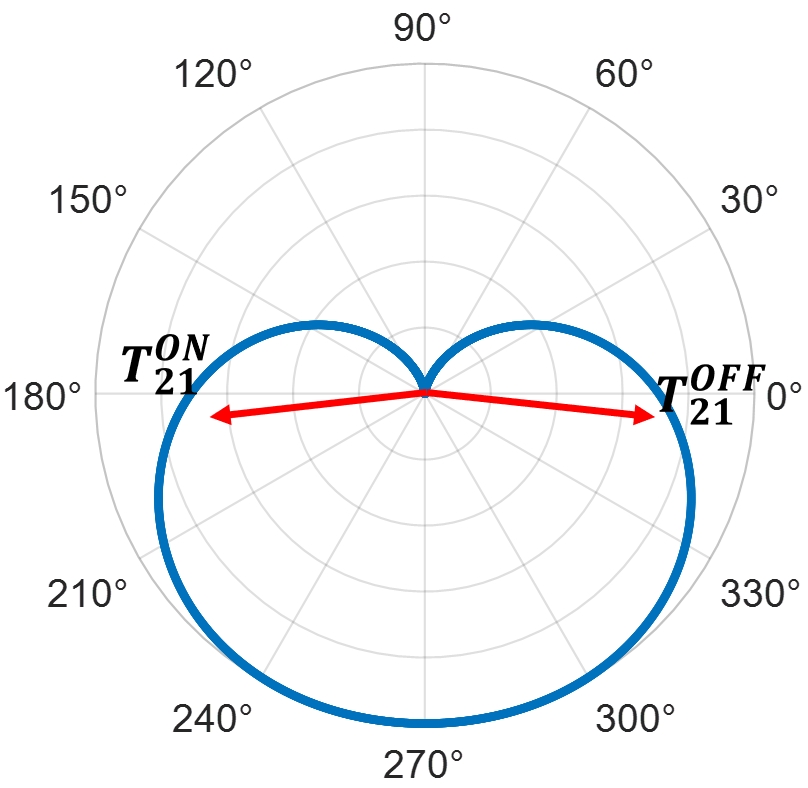} 
	} 
	\caption{Geometric illustrations of the transmission efficiency limits. (a) The transmission efficiency limit of single-switch RTA element, where the two transmission coefficients should be contained within a certain unit circle on the Smith chart. (b) The transmission efficiency limit of the cascaded double-layer RTA element, the two transmission coefficients should be contained within a certain heart-shaped curve on the Smith chart.}
	\label{chart}
\end{figure}

To provide a quantitative performance limit in general situations, this paper presents a theory on the transmission efficiency limit for single-switch RTA elements. Geometrically, this theory states that the transmission coefficients under two switch states must vary on or inside a specific unit circle with a diameter of 1 on the Smith chart. An intuitive illustration of this conclusion is shown in Fig. \ref{chart}(a). This theory reveals that the transmission phase difference is tightly constrained by the transmission amplitudes. Specifically, high transmission amplitudes result in a narrow transmission phase difference, while a phase difference of near 180$^\circ$ results in low transmission amplitudes. This conclusion is consistent with the previous finding in \cite{rta}.

Moreover, this paper expands on these insights by studying the cascaded multi-layer single-switch RTA element, as illustrated in Fig. \ref{arch}(e). When two single-switch layers are cascaded, the transmission coefficients are confined on or within specific heart-shaped curves, as illustrated in Fig. \ref{chart}(b). Therefore, through strategic designs, achieving a 1-bit phase shift with high transmission amplitudes still remains possible.

This paper is organized as follows. Section II introduces the microwave network model of the single-switch RTA elements and presents an important lemma. Section III presents the transmission efficiency limit theory for single-switch RTA elements and provides its proof. Section IV demonstrates the transmission efficiency limit theory for cascaded multi-layer RTA elements. Section V draws the conclusion.

\section{Model of Single-Switch RTA Elements}

\subsection{The Microwave Network Model of the RTA Element with a Single Switch}

\begin{figure}[!t]
	\centering
	\subfigure[] {
		\includegraphics[width=0.6\columnwidth]{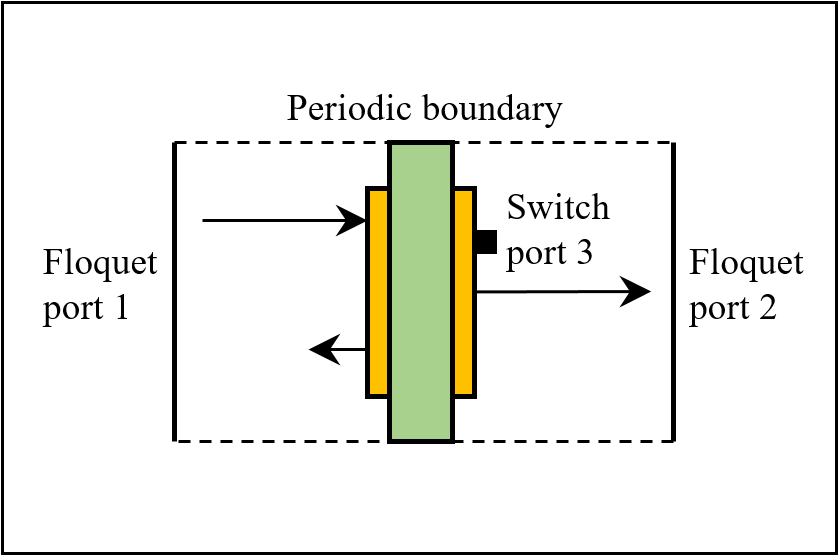} 
	} 
	\subfigure[] {
		\includegraphics[width=0.6\columnwidth]{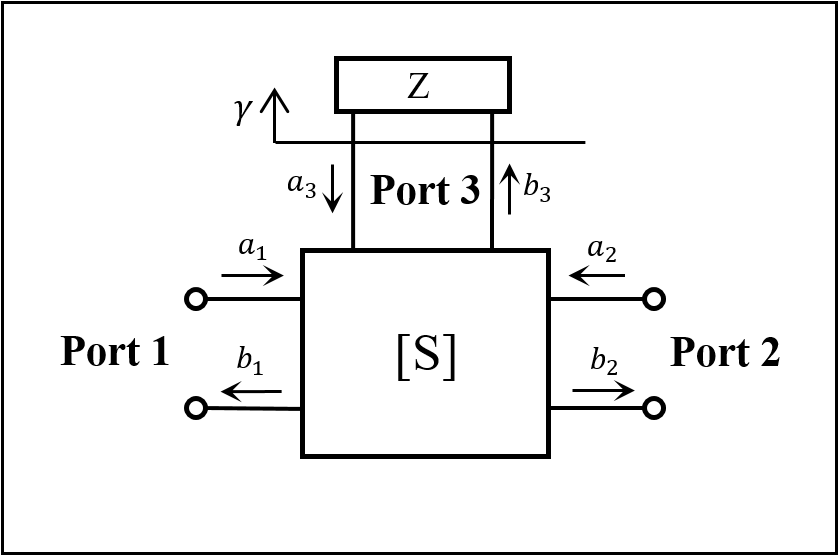} 
	} 
	\caption{The model of an RTA element with a single switch. (a) A conceptual illustration of an RTA element. (b) The equivalent network model of the RTA element.}
	\label{RTA}
\end{figure}

The RTA element with a single switch can be modeled as a three-port microwave network \cite{rta}, as shown in Fig. \ref{RTA}. The incoming wave enters the element through the Floquet port 1, and transmits toward the Floquet port 2. Since the size of the switch is much smaller than the wavelength, the lumped switch port 3 can be designated between the switch and the element. The passive structure is characterized by a 3$\times$3 scattering matrix $[S]$. According to the S-parameter definition \cite{network}, the incoming wave $[a]$ and outgoing wave $[b]$ are related through the $[S]$ matrix using the following formula
\begin{equation}
	\left[\begin{array}{l}
		b_{1} \\
		b_{2} \\
		b_{3}
	\end{array}\right]
	=
	\left[\begin{array}{lll}
		S_{11} & S_{12} & S_{13} \\
		S_{21} & S_{22} & S_{23} \\
		S_{31} & S_{32} & S_{33}
	\end{array}\right]
	\cdot
	\left[\begin{array}{l}
		a_{1} \\
		a_{2} \\
		a_{3}
	\end{array}\right].
\label{eq1}
\end{equation}
Here, $S_{ij}=S_{ji}$ due to reciprocity.

$\gamma$ is the reflection coefficient of the switch, which is defined as
\begin{equation}
	\gamma = \frac{Z-\eta}{Z+\eta} = \frac{a_3}{b_3},
\end{equation}
where $Z$ is the impedance of the switch, and $\eta$ is a reference characteristic impedance, typically 377 $\Omega$.
Notably, the switch has two states, and thus, $\gamma$ takes on two distinct values: $\gamma_\text{ON}$ and $\gamma_\text{OFF}$.

\subsection{Lemma: The Relation Among Reflection and Transmission Coefficients}

As derived in \cite{rta}, the reflection coefficient of port 1 ($\Gamma_1$) and the transmission coefficient from port 1 to port 2 ($T_{21}$) are expressed as follows 
\begin{equation}
	\Gamma_1 = \frac{b_1}{a_1} = S_{11} + \frac{S_{13}^2\gamma}{1-S_{33}\gamma},
	\label{eq2}
\end{equation}
\begin{equation}
	T_{21} =\frac{b_2}{a_1} = S_{21} + \frac{S_{23}S_{13}\gamma}{1-S_{33}\gamma}.
	\label{eq3}
\end{equation}
Similarly, the reflection coefficient of port 2 ($\Gamma_2$) is expressed as
\begin{equation}
	\Gamma_2 = \frac{b_2}{a_2} = S_{22} + \frac{S_{23}^2\gamma}{1-S_{33}\gamma}.
\end{equation}

\begin{figure}[!t]
	\centerline{\includegraphics[width=0.6\columnwidth]{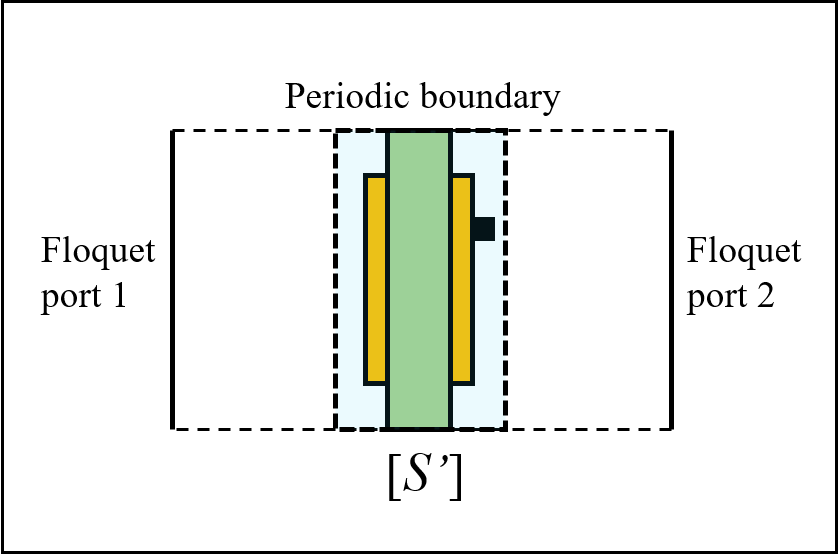}}
	\caption{The two-port microwave network model of the entire single-switch RTA element. }
	\label{s1}
\end{figure}

We denote the reflection coefficients under ON/OFF switch states as $(\Gamma^\text{ON}_1, \Gamma^\text{OFF}_1)$ and $(\Gamma^\text{ON}_2, \Gamma^\text{OFF}_2)$, and the transmission coefficients under ON/OFF switch states as $(T^\text{ON}_{21}, T^\text{OFF}_{21})$. Then we define the difference of these coefficients between two states as follows
\begin{equation}
	\Delta \Gamma_1=\Gamma^\text{ON}_1-\Gamma^\text{OFF}_1=\frac{S_{13}^2\gamma_\text{ON}}{1-S_{33}\gamma_\text{ON}}-\frac{S_{13}^2\gamma_\text{OFF}}{1-S_{33}\gamma_\text{OFF}},
	\label{eq6}
\end{equation}
\begin{equation}
	\Delta \Gamma_2=\Gamma^\text{ON}_2-\Gamma^\text{OFF}_2=\frac{S_{23}^2\gamma_\text{ON}}{1-S_{33}\gamma_\text{ON}}-\frac{S_{23}^2\gamma_\text{OFF}}{1-S_{33}\gamma_\text{OFF}},
	\label{eq7}
\end{equation}
\begin{equation}
	\Delta T_{21}=T^\text{ON}_{21}-T^\text{OFF}_{21}=\frac{S_{23}S_{13}\gamma_\text{ON}}{1-S_{33}\gamma_\text{ON}}-\frac{S_{23}S_{13}\gamma_\text{OFF}}{1-S_{33}\gamma_\text{OFF}}.
	\label{eq8}
\end{equation}
From (\ref{eq6}), (\ref{eq7}), and (\ref{eq8}), it is evident that the following relation holds
\begin{equation}
	\Delta T^2_{21} = \Delta \Gamma_1 \Delta \Gamma_2.
	\label{eq9}
\end{equation}
This expression shows a clear relationship on differences of reflection and transmission coefficients in a single-switch RTA element, and reduces the three-port network to a two-port network analysis by incorporating the switch parameters into the entire reflection and transmission coefficients, as illustrated in Fig. \ref{s1}. We define a reduced 2$\times$2 scattering matrix $[S']$ as follows
\begin{equation}
	[S']=\left[\begin{array}{ll}
		\Gamma_1 & T_{21}  \\
		T_{21} & \Gamma_2 
	\end{array}\right],
	\label{eq10}
\end{equation}
which represents the scattering behavior of the entire element.

\section{The Constraint on the Transmission Coefficients of Single-Switch RTA Elements}

In this section, we utilize (\ref{eq9}) and (\ref{eq10}) to derive the transmission efficiency limit of a single-switch RTA element. The key result is presented first, followed by proofs under both symmetric and asymmetric conditions, respectively. Additionally, the upper limit of the transmission phase difference is derived as a corollary of the theory.

\subsection{The Transmission Efficiency Limit}

The transmission efficiency limit of the single-switch RTA element is as follows: The two transmission coefficients under two switch states must lie on or within a unit circle with a diameter of 1. In other words, there exists a unit circle that can contain the two transmission coefficients for single-switch RTA element geometrically. An intuitive illustration of this finding has been shown in Fig. \ref{chart}(a). Mathematically, $T^\text{ON}_{21}$ and $T^\text{OFF}_{21}$ should satisfy the following condition
\begin{equation}
	\begin{aligned}
		\exists \varphi\in [0, 2\pi), \quad &\left|T^\text{ON}_{21}-\frac{e^{j\varphi}}{2}\right| \leq \frac{1}{2},\\
		&\left|T^\text{OFF}_{21}-\frac{e^{j\varphi}}{2}\right| \leq \frac{1}{2}.
	\end{aligned}
	\label{eq11}
\end{equation}

The proof of this statement can be divided into two parts. Firstly, for RTA elements with symmetric reflection coefficients, it can be theoretically proven that the two transmission coefficients must lie on the unit circle. Secondly, for RTA elements with asymmetric reflection coefficients, it can be observed that the two transmission coefficients must lie within the unit circle.

\subsection{Proof for Symmetric RTA Elements}

Assume the single-switch RTA element is lossless, so the scattering matrix $[S']$ is unitary, meaning that $[S'][S']^H=I$, where $H$ denotes the Hermitian transpose. Using the results of our previous derivation in \cite{rratheory}, we have
\begin{equation}
		\left \{
		\begin{array}{l}
		\left|\Gamma_1\right|^2+\left|T_{21}\right|^2=1 \\
		\left|\Gamma_2\right|^2+\left|T_{21}\right|^2=1	\\
		\end{array}
		\right.
		\label{eq13}
\end{equation}
and 
\begin{equation}
	\frac{\theta_1+\theta_2}{2}-\theta_{21} = \pm \frac{\pi}{2},
	\label{eq14}
\end{equation}
where $\theta_1$ and $\theta_2$ are the reflection phases from two ports, respectively, and $\theta_{21}$ is the transmission phase.

Here, the symmetric condition means that the microwave network is symmetric, so the two reflection coefficients are equal, i.e. $\Gamma_1=\Gamma_2$. Note that the physical structure need not be symmetric or planar, although metallic planar structures typically exhibit symmetric reflection coefficients.

\begin{figure}[!t]
	\centering
	\subfigure[] {
		\includegraphics[width=0.45\columnwidth]{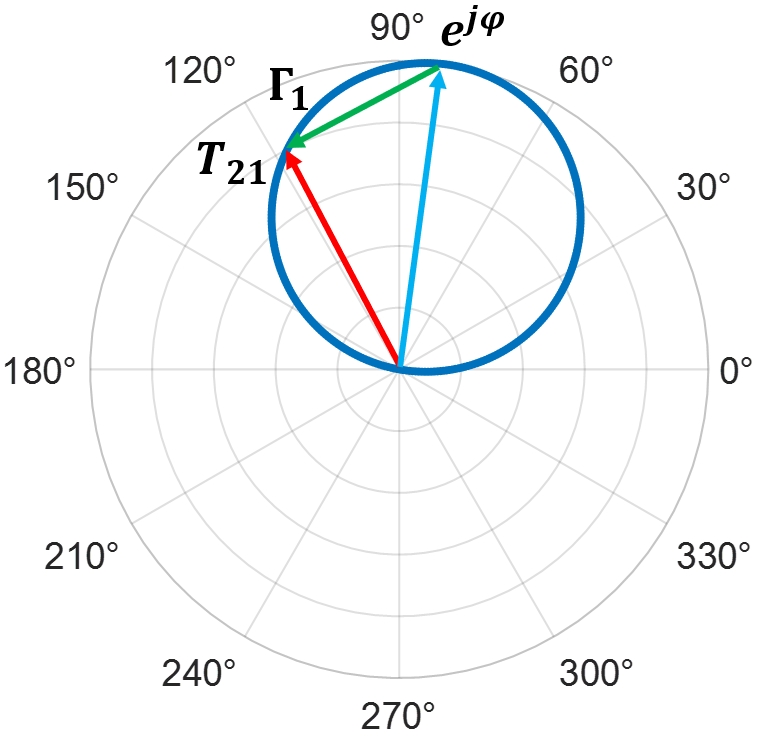} 
	} 
	\subfigure[] {
	\includegraphics[width=0.45\columnwidth]{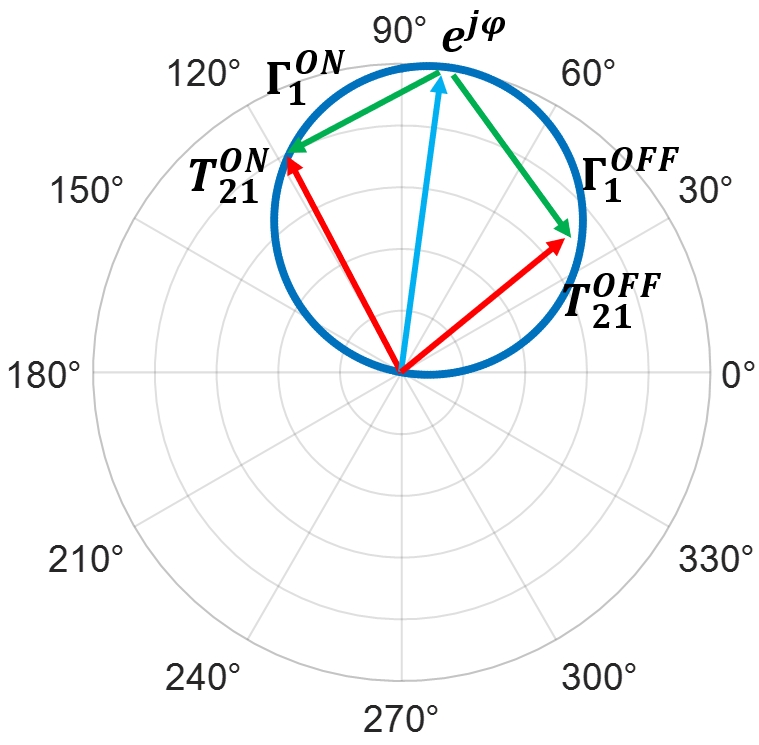} 
} 
	\caption{The geometric illustration of the relations between $T_{21}$ and $\Gamma_1$ on the Smith chart for the symmetric case. (a) An illustration of the relation in (\ref{eq17}). (b) Illustration of the geometric method to solve (\ref{eq19}). $T^\text{ON}_{21}$ and $T^\text{OFF}_{21}$ vary on the same unit circle in case $\Delta T_{21} = \Delta \Gamma_1$.}
	\label{sym1}
\end{figure}

Given $\Gamma_1 = \Gamma_2$, it follows that $\theta_1 = \theta_2$. Combining these results with (\ref{eq13}) and (\ref{eq14}), the equations simplify to:
\begin{equation}
	\left \{
	\begin{array}{l}
		\left|\Gamma_1\right|^2+\left|T_{21}\right|^2=1 \\
		\theta_1-\theta_{21} = \pm \frac{\pi}{2}	\\
	\end{array}
	\right.
	\label{eq16}
\end{equation}
It can be mathematically shown that the complex numbers $\Gamma_1$ and $T_{21}$ have the following relations
\begin{equation}
	T_{21}-\Gamma_1=e^{j\varphi},
	\label{eq17}
\end{equation}
where $\varphi$ is a parameter of this equation.
Geometrically, as shown in Fig. \ref{sym1}(a), vectors $\Gamma_1$ and $T_{21}$ on the complex plane are perpendicular according to the Pythagorean theorem, with the hypotenuse length fixed at 1. Therefore, $T_{21}$, $e^{j\varphi}$ and $0$ form a circle on the Smith chart, with $e^{j\varphi}$ as the diameter, which can be used to characterize the circle. Consequently, $T_{21}$ varies along a circle defined by $e^{j\varphi}$ on the Smith chart.

Since $\Gamma_1=\Gamma_2$ under both switch states, the difference between two states should also be equal, i.e., $\Delta\Gamma_1=\Delta\Gamma_2$. Substituting this relation into (\ref{eq9}), the following expression is obtained
\begin{equation}
	\Delta T^2_{21} = \Delta \Gamma^2_1.
	\label{eq19}
\end{equation}
This equation has two sets of solutions, and this discussion will focus on the case where $\Delta T_{21} = \Delta \Gamma_1$. The analysis for $\Delta T_{21} = -\Delta \Gamma_1$ is similar. It is assumed that $T^\text{ON}_{21}$ varies on the circle determined by $e^{j\varphi^\text{ON}}$, and $T^\text{OFF}_{21}$ varies on the circle determined by $e^{j\varphi^\text{OFF}}$.

Since $\Delta T_{21} = \Delta \Gamma_1$, it follows that
\begin{equation}
	T^\text{ON}_{21}-\Gamma^\text{ON}_{1}=T^\text{OFF}_{21}-\Gamma^\text{OFF}_{1}.
	\label{eq20}
\end{equation} 
Based on (\ref{eq17}) and (\ref{eq20}), it is found that
\begin{equation}
	e^{j\varphi^\text{ON}}=e^{j\varphi^\text{OFF}},
\end{equation}
which means that $T^\text{ON}_{21}$ and $T^\text{OFF}_{21}$ vary on the same unit circle, as shown in Fig. \ref{sym1}(b). Therefore, for lossless symmetric RTA elements, both transmission coefficients should lie on one unit circle.

\subsection{Simulation Demonstration of Symmetric Single-Switch RTA Elements}

\begin{figure}[!t]
	\centerline{\includegraphics[width=0.6\columnwidth]{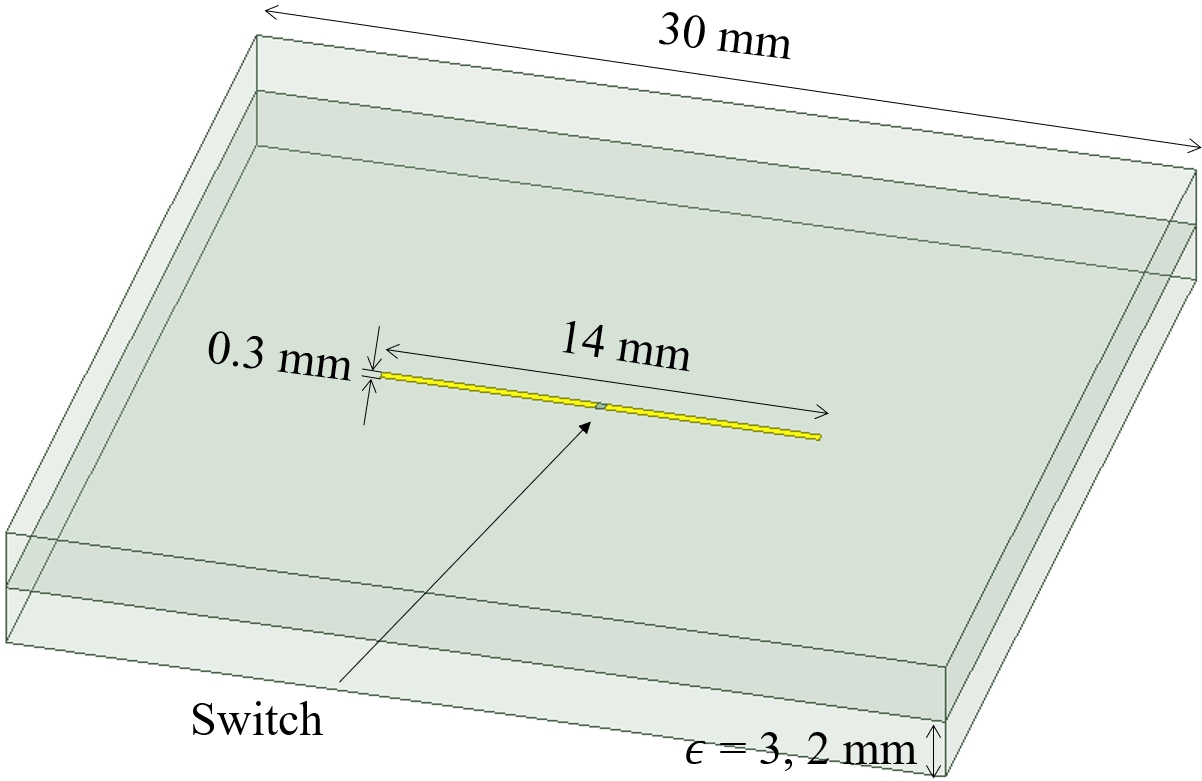}}
	\caption{Schematic of the symmetric sandwiched dipole RTA element model with a single switch used for simulation verification in the symmetric case.}
	\label{symmodel}
\end{figure}

To confirm the conclusions regarding general symmetric networks, a sandwiched single-switch dipole RTA element is modeled and simulated using Ansys HFSS 2020, as shown in Fig. \ref{symmodel}. For simplicity, the dipole is assigned as a perfect electric conductor (PEC), sandwiched by two identical substrates with thickness of 2 mm and a permittivity of 3. This configuration ensures that the structure is symmetric, thus the reflection coefficients are symmetric. The lumped switch in the model is represented by a variable capacitance ranging from 1 fF to 1000 fF, enabling the generation of diverse transmission coefficients. The simulation frequency is 6.37 GHz.

\begin{figure}[!t]
	\centerline{\includegraphics[width=0.5\columnwidth]{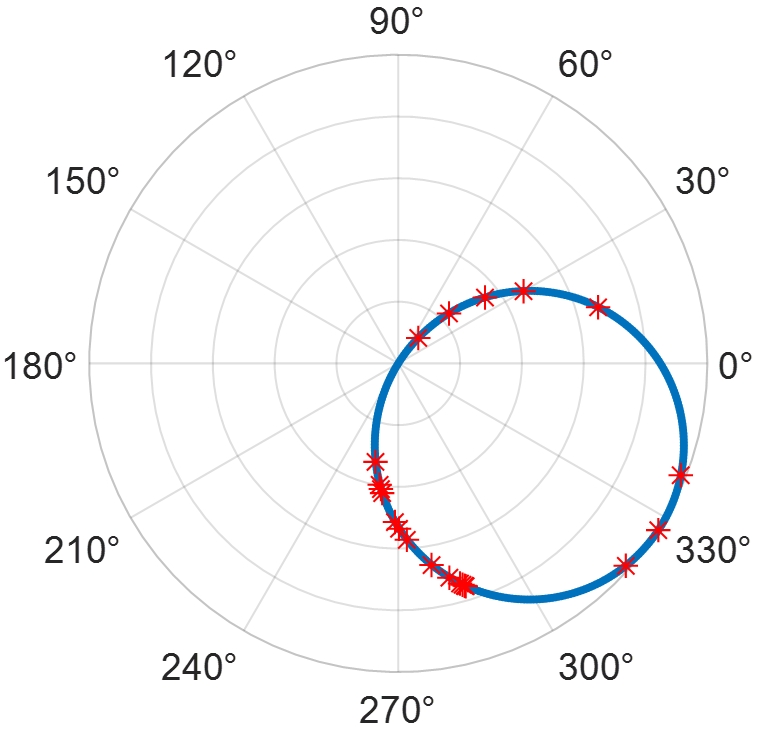}}
	\caption{The transmission coefficients of the element with different switch parameters (red star). All these transmission coefficients lie on a unit circle (blue circle).}
	\label{symcoeff}
\end{figure}

Fig. \ref{symcoeff} illustrates the transmission coefficients for different switch parameters. Consistent with theoretical predictions, the transmission coefficients across the range of capacitance values trace a circle on the Smith chart. The rotation angle of this circle is influenced by the substrate thickness. In summary, the simulation example shows that regardless of the switch states, all the transmission coefficients of a symmetric RTA element are constrained by a circle at given structures.

\subsection{Analysis of Asymmetric RTA Elements}

For asymmetric structures, it is necessary to have $\Gamma_1\neq\Gamma_2$ under at lease one state. This part analyzes the asymmetric single-switch RTA elements by numerical calculation, and it is observed that the two transmission coefficients should vary inside the unit circle.

Assume the element is lossless, so (\ref{eq13}) and (\ref{eq14}) are still applicable. From (\ref{eq13}), it can be inferred that $\left|\Gamma_1\right|=\left|\Gamma_2\right|$. Considering only the relative relationship between two transmission coefficients can influence the performance of the RTA element, one of the transmission coefficients, such as $T^\text{ON}_{21}$, can be assumed to vary on the unit circle determined by $e^{j0} = 1$. Thus, $T^\text{ON}_{21}$ can be expressed as
\begin{equation}
T^\text{ON}_{21}= \left|\cos(\theta^\text{ON}_{21})\right|e^{j\theta^\text{ON}_{21}}
\label{eq21}
\end{equation}

Assuming that the transmission coefficient in the OFF state varies on a circle determined by $e^{j\varphi}$, where $\varphi$ is an independent variable, $T^\text{OFF}_{21}$ can be expressed as
\begin{equation}
T^\text{OFF}_{21}=\left|\cos(\theta^\text{OFF}_{21}-\varphi)\right|e^{j\theta^\text{OFF}_{21}}
\label{eq22}
\end{equation}

\begin{figure}[!t]
	\centering
	\subfigure[] {
		\includegraphics[width=0.46\columnwidth]{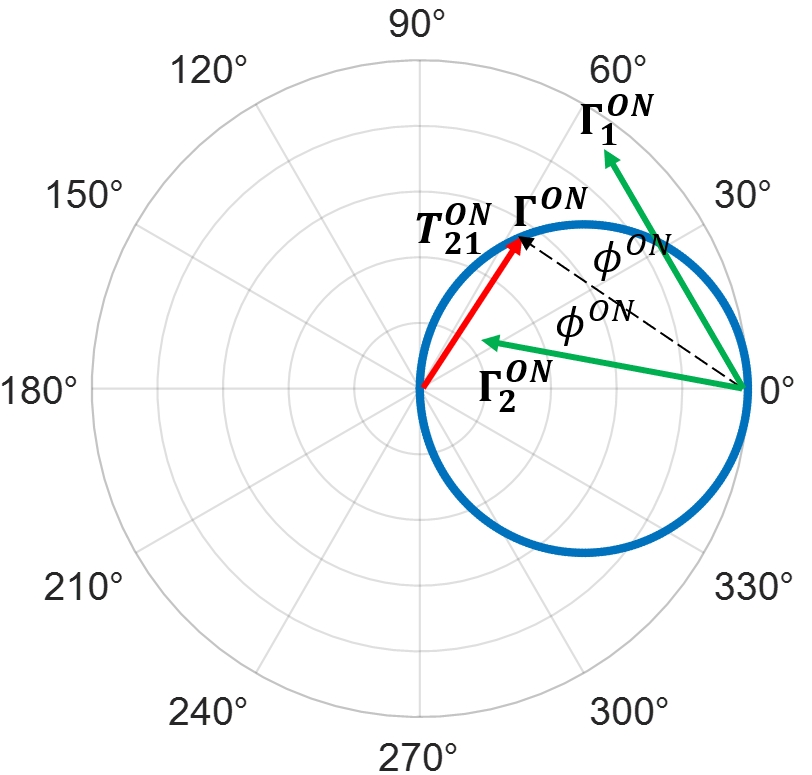} 
	} 
	\subfigure[] {
		\includegraphics[width=0.46\columnwidth]{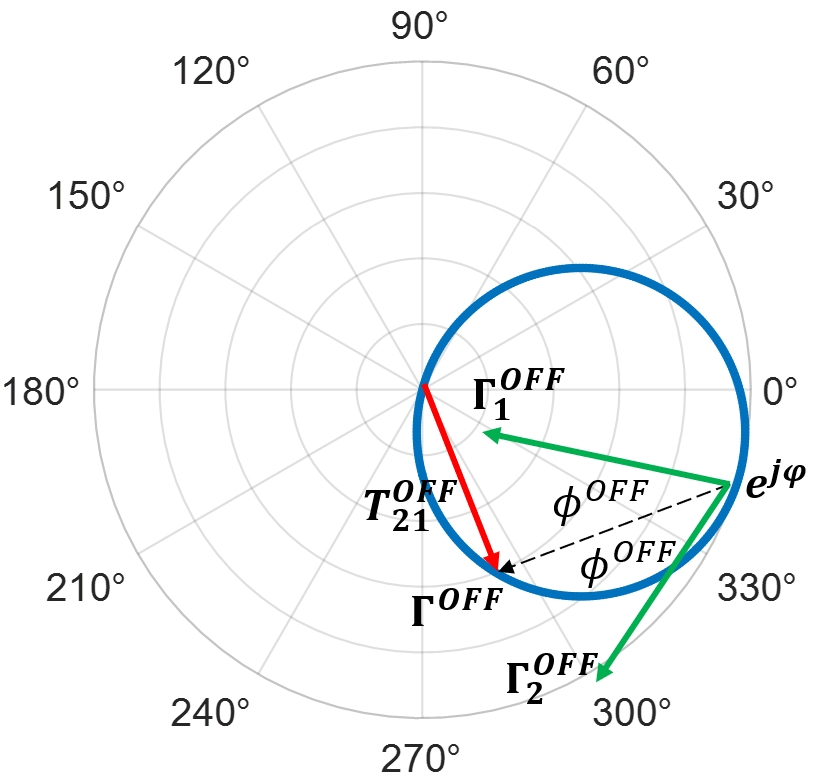} 
	} 
	\caption{Geometric illustration of the relation among reflection and transmission coefficients given by (\ref{eq13}) and (\ref{eq14}) in asymmetric case. (a) The relations at ON state. (b) The relations at OFF state.}
	\label{asym}
\end{figure}

An auxiliary variable $\Gamma^\text{ON}$ is introduced, with the equal amplitude to $\Gamma^\text{ON}_1$ and $\Gamma^\text{ON}_2$, and the phase is defined as $\theta^\text{ON} = (\theta^\text{ON}_1+\theta^\text{ON}_2)/2$. Therefore, the relation between $\Gamma^\text{ON}$ and $T^\text{ON}_{21}$ can be expressed as
\begin{equation}
	\left \{
	\begin{array}{l}
		\left|\Gamma^\text{ON}\right|^2+\left|T^\text{ON}_{21}\right|^2=1 \\
		\theta^\text{ON}-\theta^\text{ON}_{21} = \pm \frac{\pi}{2}	\\
	\end{array}
	\right.
	\label{eq24}
\end{equation}
Similar to the analysis in the symmetric case, $\Gamma^\text{ON}$ should be perpendicular to $T^\text{ON}_{21}$, and the length of the hypotenuse is 1. Fig. \ref{asym}(a) shows a graphical representation of the relationship between reflection and transmission coefficients.

Additionally, since $\theta^\text{ON} = (\theta^\text{ON}_1 + \theta^\text{ON}_2) / 2$, $\Gamma^\text{ON}$ is positioned on the angular bisector between $\Gamma^\text{ON}_1$ and $\Gamma^\text{ON}_2$. Denoting the phase difference between $\Gamma^\text{ON}$ and $\Gamma^\text{ON}_1$ as $\phi^\text{ON}$, the phase difference with $\Gamma^\text{ON}_2$ will be $-\phi^\text{ON}$. From these geometric relationships, the reflection coefficients are described as
\begin{equation}
	\left \{
	\begin{array}{l}
		\Gamma^\text{ON}_1=(T^\text{ON}_{21}-1)e^{-j\phi^\text{ON}} \\
		\Gamma^\text{ON}_2=(T^\text{ON}_{21}-1)e^{j\phi^\text{ON}}	\\
	\end{array}
	\right.
	\label{eq25}
\end{equation}

\begin{figure}[!t]
	\centering
	\subfigure[] {
		\includegraphics[width=0.45\columnwidth]{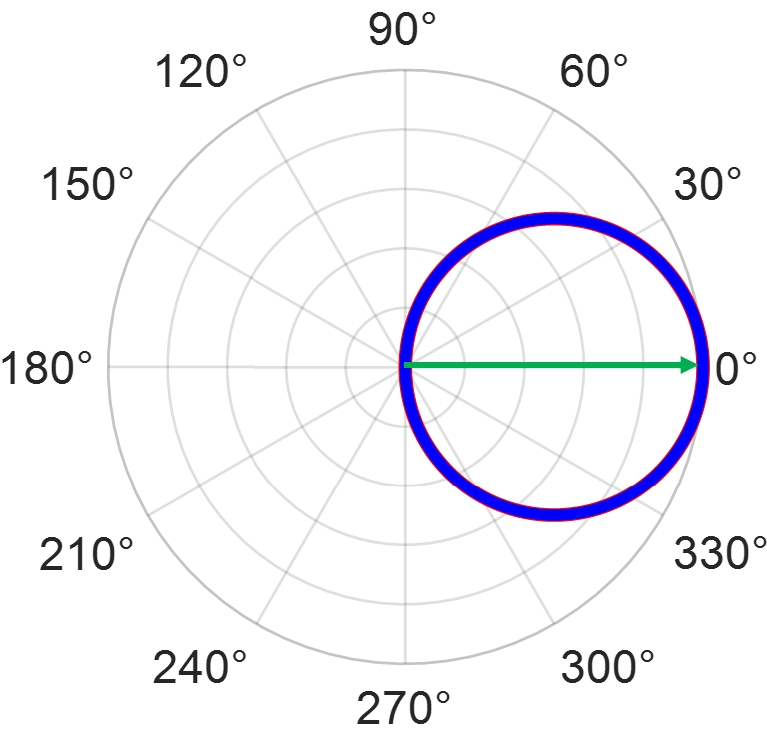} 
	} 
	\subfigure[] {
		\includegraphics[width=0.45\columnwidth]{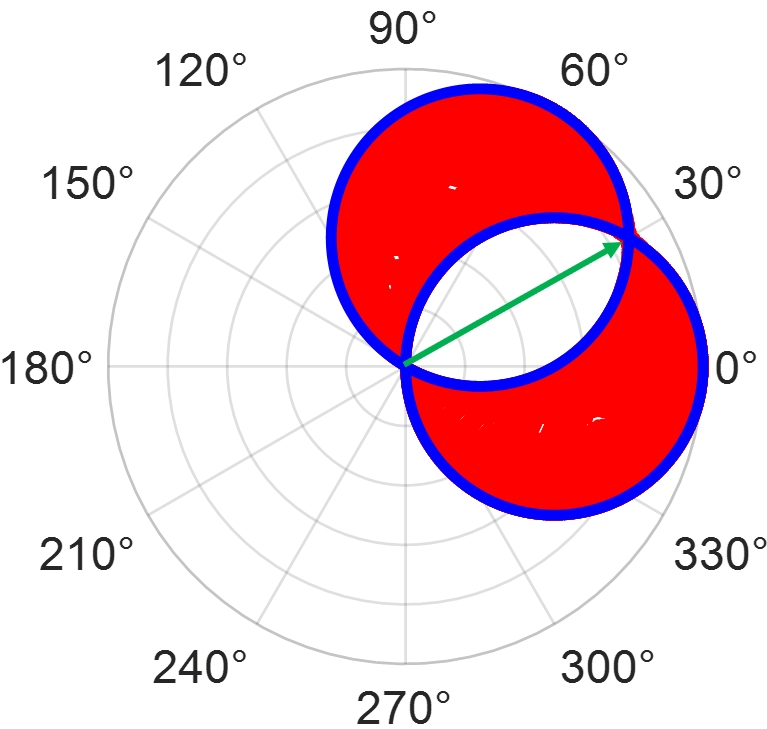} 
	} 
	\subfigure[] {
		\includegraphics[width=0.45\columnwidth]{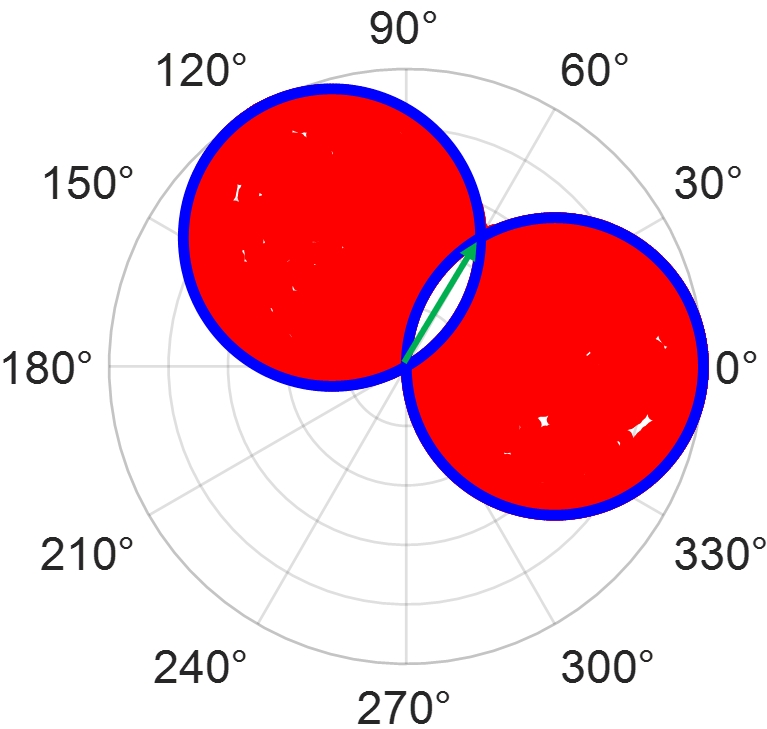} 
	} 
	\subfigure[] {
		\includegraphics[width=0.45\columnwidth]{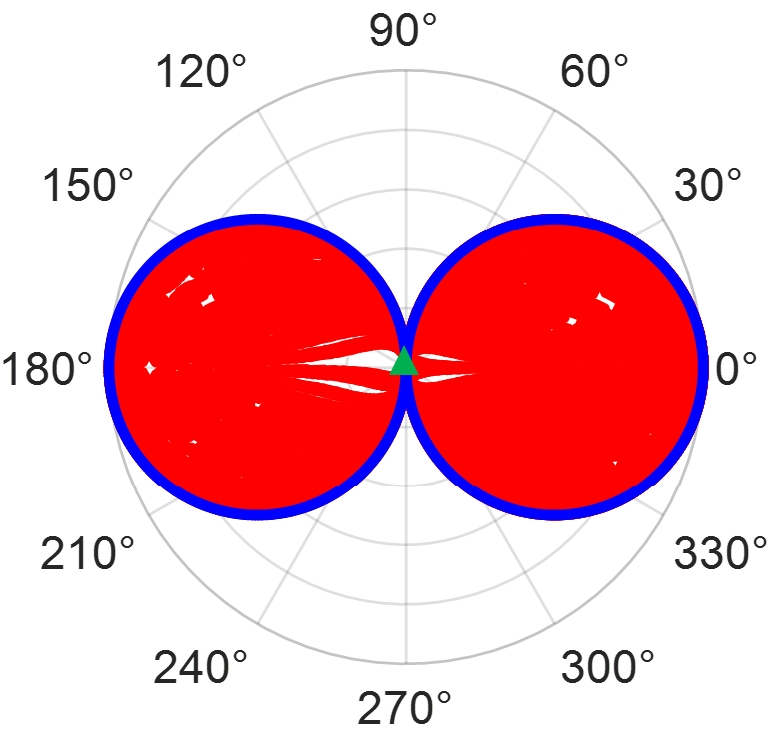} 
	} 
	\caption{An illustration of the relative relation between two transmission coefficients in asymmetric case. Green arrow: $T^\text{ON}_{21}$; blue circles: the circles on which $T^\text{ON}_{21}$ lies; red region: the variation range of $T^\text{OFF}_{21}$. 	(a) $\theta^\text{ON}_{21}=0$. (b) $\theta^\text{ON}_{21}=\frac{\pi}{6}$. (c) $\theta^\text{ON}_{21}=\frac{\pi}{3}$. (d) $\theta^\text{ON}_{21}=\frac{\pi}{2}$.}
	\label{asymnum}
\end{figure}

Similarly, as shown in Fig. \ref{asym}(b), the reflection coefficients at OFF state are expressed as
\begin{equation}
	\left \{
	\begin{array}{l}
		\Gamma^\text{OFF}_1=(T^\text{OFF}_{21}-e^{j\varphi})e^{-j\phi^\text{OFF}} \\
		\Gamma^\text{OFF}_2=(T^\text{OFF}_{21}-e^{j\varphi})e^{j\phi^\text{OFF}}	\\
	\end{array}
	\right.
	\label{eq26}
\end{equation}

By substituting (\ref{eq25}) and (\ref{eq26}) into (\ref{eq9}), the expression is finally simplified as
\begin{equation}
	\begin{aligned}
		\left(T^\text{ON}_{21}-T^\text{OFF}_{21}\right)^2&=\left(T^\text{ON}_{21}-1\right)^2+\left(T^\text{OFF}_{21}-e^{j\varphi}\right)^2\\
		-&2\cos(\Delta\phi)\left(T^\text{ON}_{21}-1\right)\left(T^\text{OFF}_{21}-e^{j\varphi}\right),
	\end{aligned}
\label{eq27}
\end{equation}
where $\Delta\phi=\phi^\text{OFF}-\phi^\text{ON}$ denotes difference between the two-state asymmetric reflection phases. $\Delta\phi$ characterizes the extent of asymmetry. Note that (\ref{eq27}) involves four parameters: $\theta^\text{ON}_{21}$, $\theta^\text{OFF}_{21}$, $\varphi$ and $\Delta\phi$, with consideration to (\ref{eq21}) and (\ref{eq22}).

Theoretically, given $\theta^\text{ON}_{21}$, $\varphi$, and $\Delta\phi$, $\theta^\text{OFF}_{21}$ can be solved using (\ref{eq27}). Subsequently, the mathematical criterion in (\ref{eq11}) can determine whether $T^\text{OFF}_{21}$ and $T^\text{ON}_{21}$ lie within the same circle. However, the expression for $\theta^\text{OFF}_{21}$ is very complex and not intuitive. Therefore, numerical calculations are used to demonstrate the relation between $T^\text{OFF}_{21}$ and $T^\text{ON}_{21}$.

By fixing $\theta^\text{ON}_{21}$, all possible values of $\theta^\text{OFF}_{21}$, $\varphi$, and $\Delta\phi$ are numerically enumerated. The combinations that satisfy (\ref{eq27}) are selected, and the corresponding values for $T^\text{OFF}_{21}$ are plotted on the Smith chart. For illustration, several examples are presented, demonstrating that regardless of $\theta^\text{ON}_{21}$ variations, $T^\text{OFF}_{21}$ always resides within the circle on which $T^\text{ON}_{21}$ is located, as shown in Fig. \ref{asymnum}. It can be concluded by the case studies that for an asymmetric RTA element, $T^\text{OFF}_{21}$ and $T^\text{ON}_{21}$ are always located inside the same circle. Furthermore, it is evident from these observations that asymmetric RTA elements exhibit comparatively weaker performance than symmetric RTA elements. Therefore, the unit circle on which the symmetric elements lie presents the general performance limit of single-switch RTA elements.

\subsection{Simulation Demonstration of Asymmetric Single-Switch RTA Elements}

\begin{figure}[!t]
	\centerline{\includegraphics[width=0.7\columnwidth]{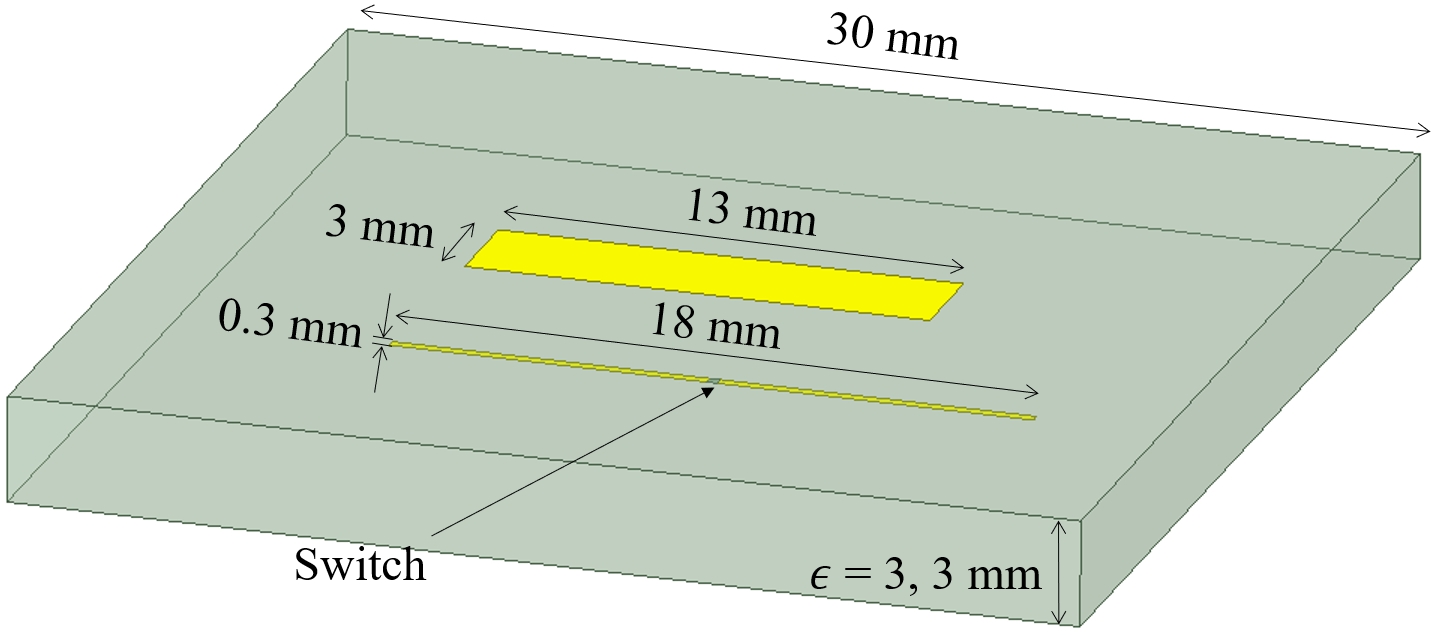}}
	\caption{Simulation model of an asymmetric RTA element with two asymmetric dipoles. The top dipole is pure metal, and the bottom one is loaded with tunable switches.}
	\label{asymmodel}
\end{figure}

To demonstrate the impact of asymmetry, another metal dipole is positioned above the diode-tuned dipole, breaking the symmetric reflection condition from two Floquet ports, as shown in Fig. \ref{asymmodel}. The simulation is conducted in Ansys HFSS 2020, with metal assigned as a PEC. An ideal switch with no resistance and a capacitance range of 1 fF to 1000 fF is utilized. The simulation frequency is set at 7.1 GHz.

\begin{figure}[!t]
	\centerline{\includegraphics[width=0.5\columnwidth]{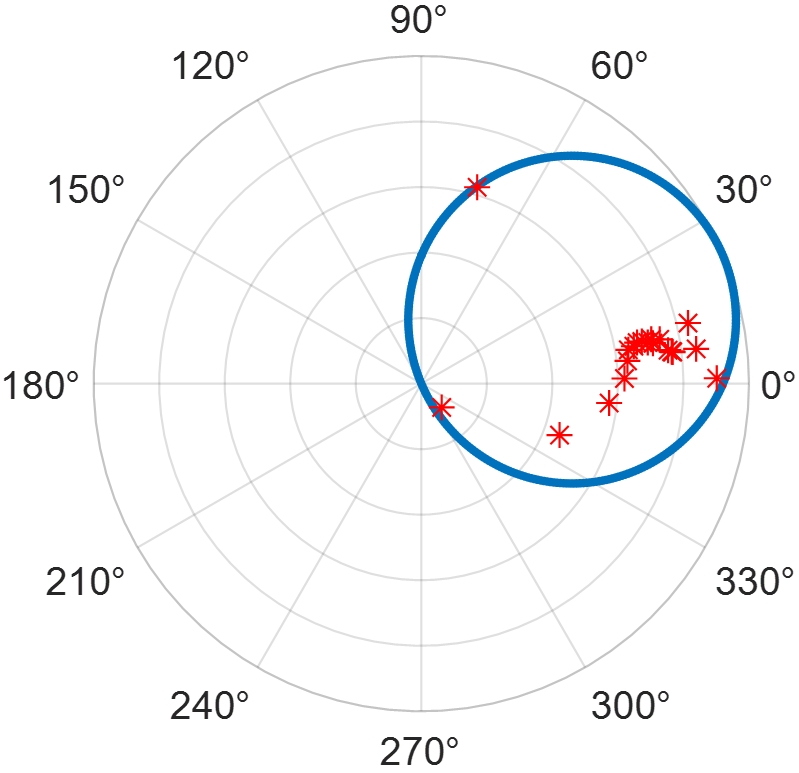}}
	\caption{The transmission coefficients of the asymmetric element with difference switch states (red star). All these transmission coefficients lie within a unit circle (blue circle).}
	\label{asymcoeff}
\end{figure}

Fig. \ref{asymcoeff} illustrates the transmission coefficients of the asymmetric RTA element across all switch states. It is observed that there is a specific circle that covers all the transmission coefficients under different switch states. Therefore, for this particular structure, all the transmission coefficients of the asymmetric structures are constrained within one unit circle.

\subsection{Inference: The Transmission Phase Difference Limit of Single-Switch RTA Elements}

Based on the findings above, it is useful to derive the constraint on the transmission phase difference and transmission amplitudes. 

\begin{figure}[!t]
	\centerline{\includegraphics[width=0.5\columnwidth]{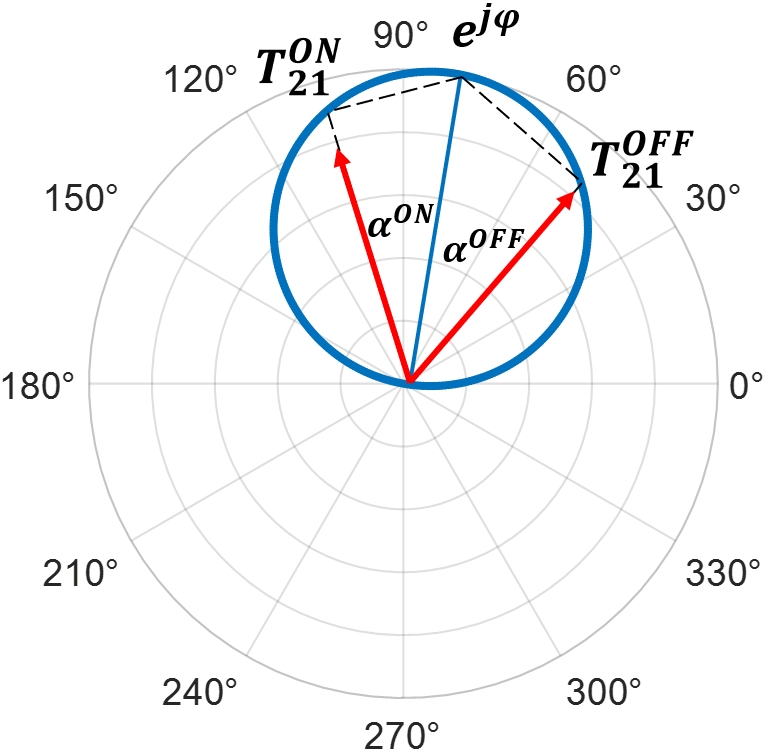}}
	\caption{Geometric illustration of the relationship among transmission phases and transmission amplitudes.}
	\label{phaseamp}
\end{figure}

As shown in Fig. \ref{phaseamp}, the transmission phase difference is defined as the included angle between the two transmission coefficients. Since both two transmission coefficients lie on or inside a unit circle, the relative phases of these transmission coefficients can be expressed as
\begin{equation}
	\left \{
	\begin{array}{l}
		\cos(\alpha^\text{ON})\geq \left|T^\text{ON}_{21}\right|\\
		\cos(\alpha^\text{OFF})\geq \left|T^\text{OFF}_{21}\right|	\\
	\end{array}
	\right.
	\label{eq29}
\end{equation}
where $\alpha$ represents the included angle between one transmission coefficient and the diameter. Given the transmission amplitudes, the phase difference $\Delta\theta$ between the two transmission coefficients is limited by
\begin{equation}
\Delta\theta \leq \alpha^\text{ON}+\alpha^\text{OFF}\leq\arccos(\left|T^\text{ON}_{21}\right|)+\arccos(\left|T^\text{OFF}_{21}\right|).
\label{eq30}
\end{equation}
The equal sign holds only when $T^\text{ON}_{21}$ and $T^\text{OFF}_{21}$ both lie on the circle, and are positioned on opposite sides of the diameter.

\begin{figure}[!t]
	\centering
	\subfigure[] {
		\includegraphics[width=0.55\columnwidth]{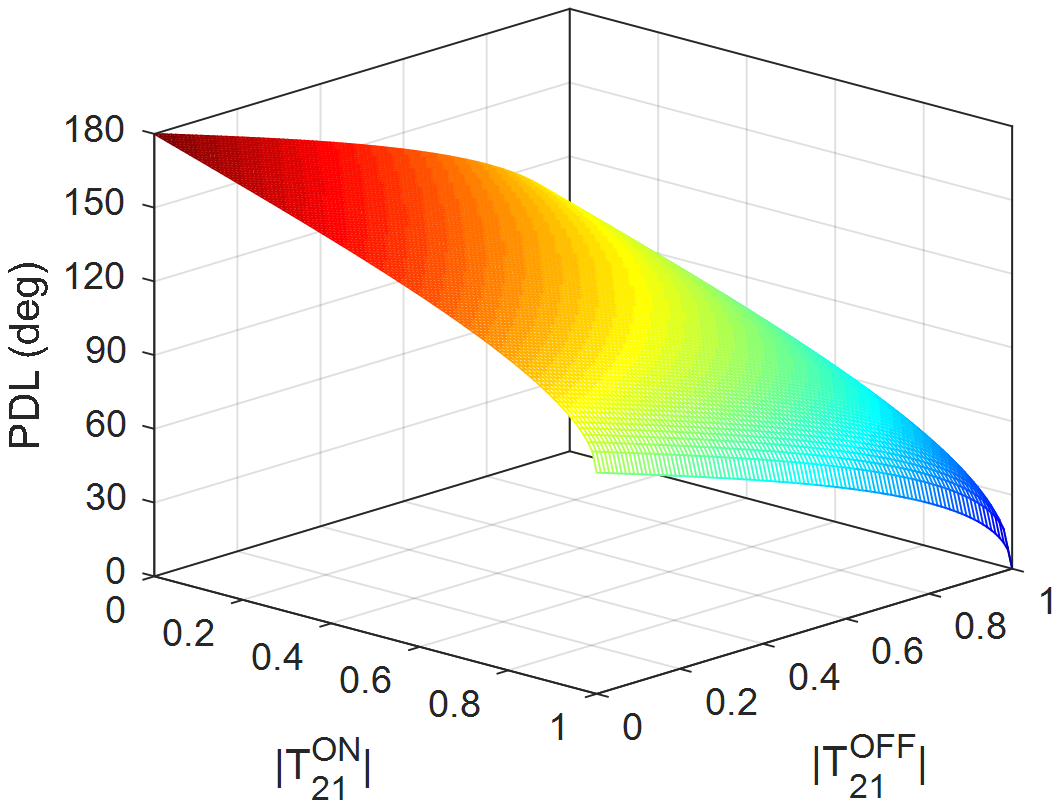} 
	} 
	\subfigure[] {
		\includegraphics[width=0.55\columnwidth]{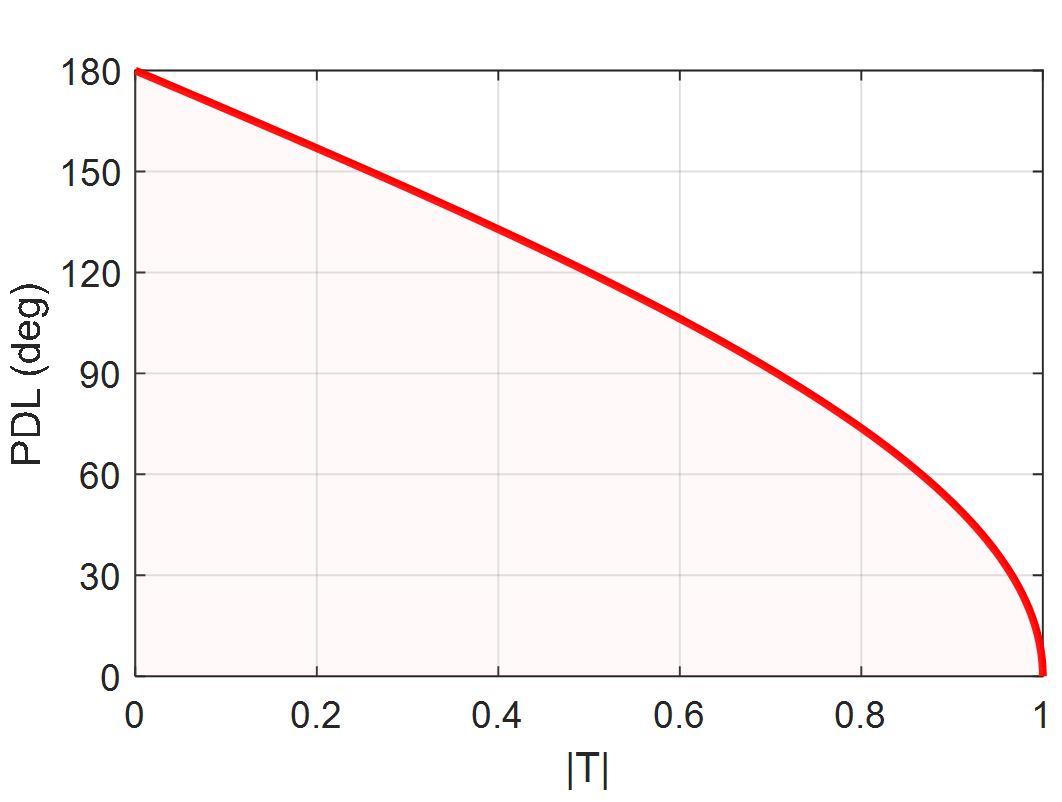} 
	} 
	\caption{The relationship among transmission phase difference limit and transmission amplitudes. (a) The 3-D view. (b) The 2-D view when the transmission amplitudes are equal. The feasible region is represented by the area below the curve.}
	\label{PDL}
\end{figure}

The maximum transmission phase difference is limited by the two transmission amplitudes, as expressed in (\ref{eq30}). The phase difference limit (PDL) is defined as the sum of the two arccosine terms, $\arccos(\left|T^\text{ON}_{21}\right|)+\arccos(\left|T^\text{OFF}_{21}\right|)$. The relationship among PDL and transmission amplitudes is plotted in Fig. \ref{PDL}(a). Particularly, when the transmission amplitudes are equal, i.e.,  $\left|T\right|=\left|T^\text{ON}_{21}\right|=\left|T^\text{OFF}_{21}\right|$, and the relationship between PDL and $\left|T\right|$ is plotted in Fig. \ref{PDL}(b). 

These curves in Fig. \ref{PDL} reveal that a single-switch RTA element cannot achieve both large phase difference and high transmission amplitudes simultaneously. For instance, to achieve high transmission amplitudes close to 1, the phase difference must be near 0. Alternatively, a phase difference of approximately 180$^\circ$ requires the transmission amplitudes to approach zero. Moreover, achieving a 90$^\circ$ phase difference introduces at least a 3-dB amplitude loss. These results quantitatively support the findings in \cite{rta} that it is impossible to achieve high transmission efficiency with a large phase tuning range using a single-switch RTA element.

\section{Transmission Efficiency Limit of Cascaded Double-Layer RTA Elements and Beyond}

\subsection{Theoretical Analysis of Cascaded Symmetric RTA Elements with the Same Structures}

\begin{figure}[!t]
	\centerline{\includegraphics[width=0.5\columnwidth]{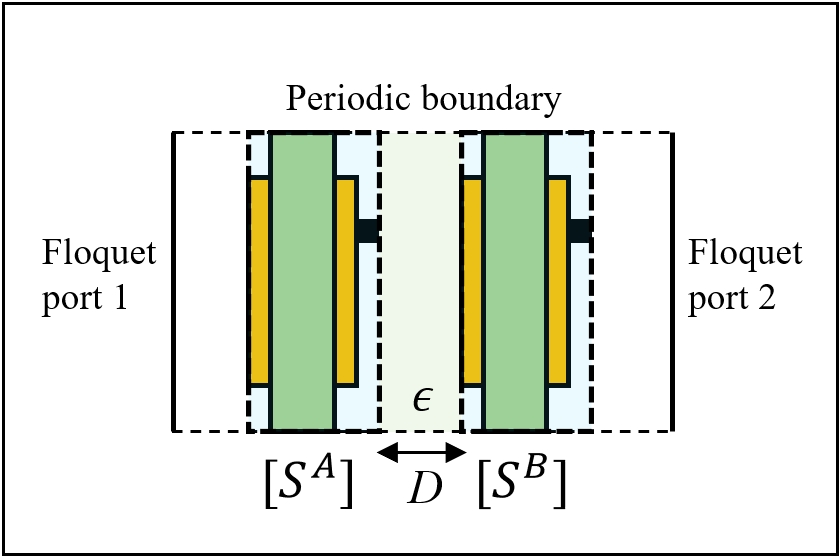}}
	\caption{Schematic of the cascaded double-layer RTA element.}
	\label{cascadesche}
\end{figure}

The schematic of the cascaded double-layer structure is shown in Fig. \ref{cascadesche}. Two layers are designated as layer A and layer B, spaced at a distance $D$. In practice, these layers are mounted on a substrate with permittivity $\epsilon$. 

\begin{figure}[!t]
	\centerline{\includegraphics[width=0.8\columnwidth]{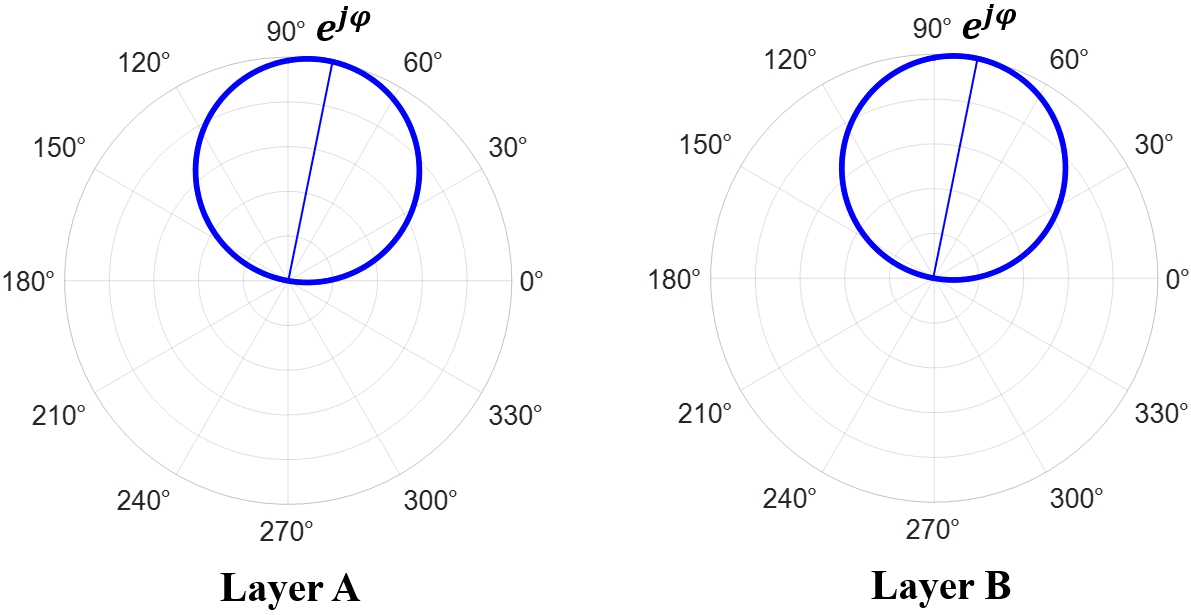}}
	\caption{The varying range for transmission coefficients of two single symmetric layers with identical structures and states.}
	\label{layerab}
\end{figure}

The analysis begins by considering the scenario where the cascaded double-layer RTA element possesses identical symmetric structures, and the switches in both layers change synchronously, thus $[S^A]$ and $[S^B]$ are equal. Following the results outlined in Section III.B, the transmission coefficients of both layers should vary on the same unit circles defined by $\varphi$, as illustrated in Fig. \ref{layerab}.

\begin{figure}[!t]
		\centering
	\subfigure[] {
	\includegraphics[width=0.6\columnwidth]{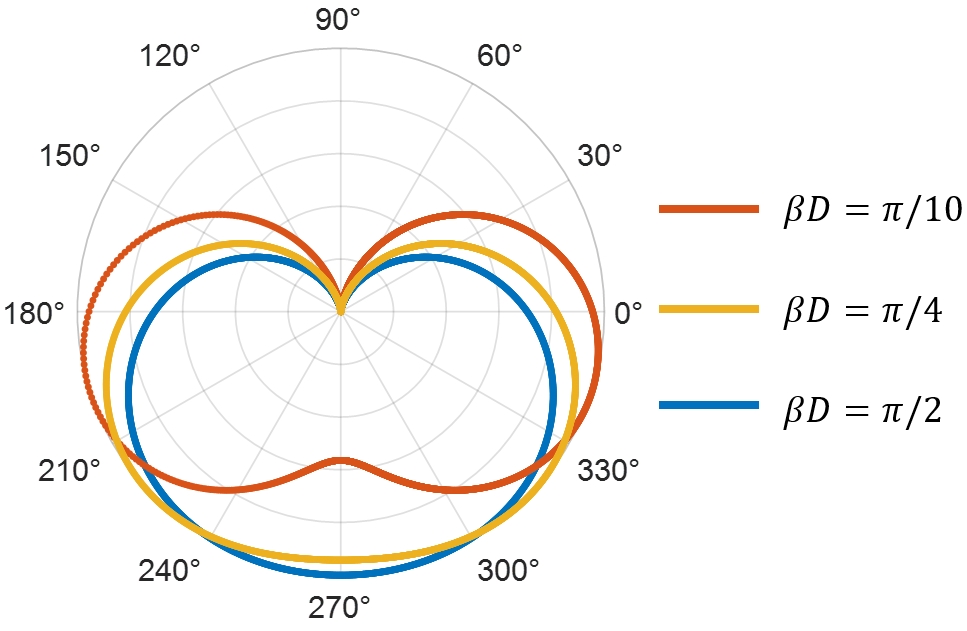} 
} 
\subfigure[] {
	\includegraphics[width=0.6\columnwidth]{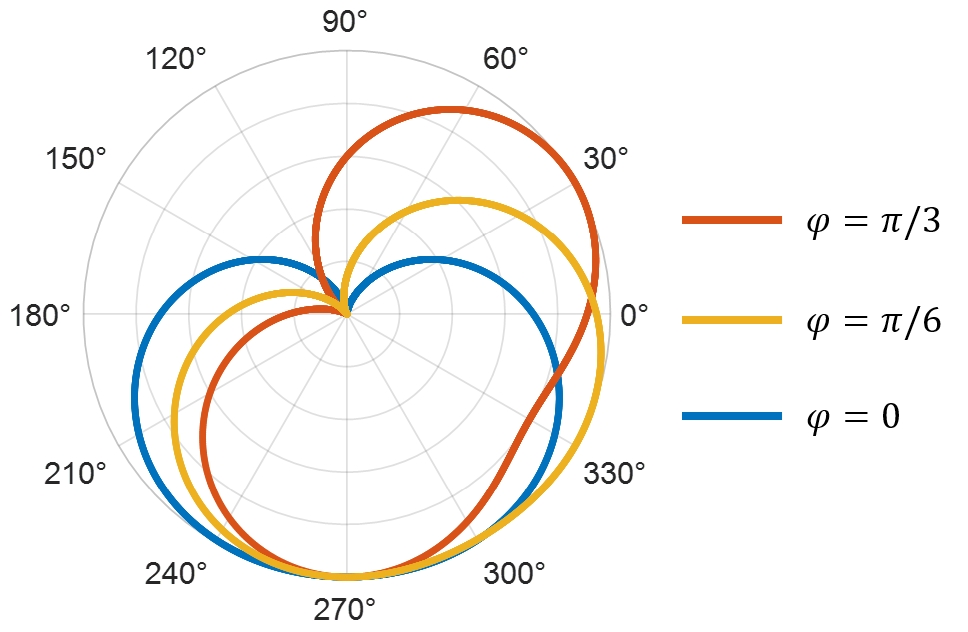} 
} 
	\caption{Transmission coefficients of the double-layer RTA element with the same symmetric structures. (a) Different substrate thickness with dielectric permittivity of $\epsilon$ = 1 and $\varphi$ = 0. (b) Different circle parameters $\varphi$ with dielectric permittivity of $\epsilon$ = 1 and thickness of $\beta D$ = $\pi$/2.}
	\label{dualsym}
\end{figure}

To obtain the possible transmission coefficients for the cascaded structure, the cascading formula of the S matrix is applied, as detailed in the appendix of \cite{FSS}. The analytical results are shown in Fig. \ref{dualsym}. Initially, different substrate thicknesses $D$ are considered, fixing the $\epsilon$ = 1 and $\varphi$ = 0, as shown in Fig. \ref{dualsym}(a). Here, $\beta=2\pi\sqrt{\epsilon}/\lambda$. The transmission coefficients trace a set of heart-shaped curves. As the thickness decreases, the curve flattens, indicating the potential for achieving 1-bit phase modulation with high transmission amplitude. Subsequently, the thickness is fixed as $\beta D$ = $\pi$/2 and the permittivity of $\epsilon$ = 1, while the effects of varying unit circles for each single layer are explored. As shown in Fig. \ref{dualsym}(b), as $\varphi$ increases, the curve not only rotates but also becomes flatter, also suggesting the feasibility of high-performance 1-bit transmission.

In summary, for cascaded double-layer RTA elements with identical symmetric structure and switch configurations, the varying range of transmission coefficients are expanded. They are constrained on a set of heart-shaped curves, thus certain switch states enable a 1-bit phase shift with high transmission amplitude.

\subsection{Simulation Demonstration of Cascaded Symmetric RTA Elements}

\begin{figure}[!t]
	\centerline{\includegraphics[width=0.5\columnwidth]{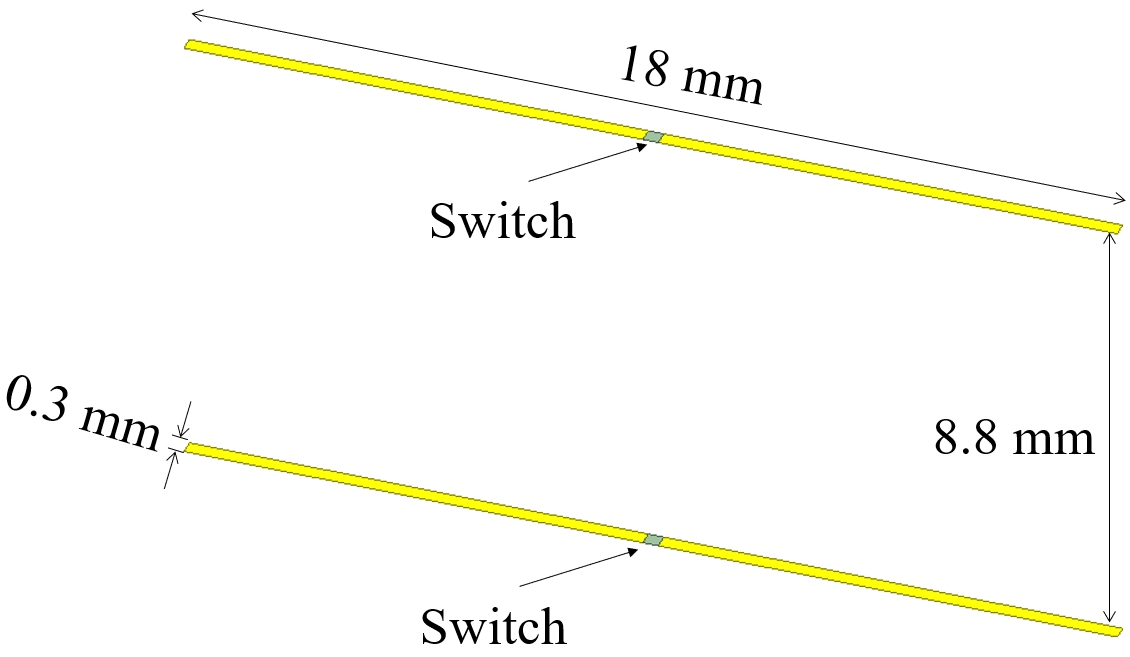}}
	\caption{Simulation model of the double-layer symmetric elements with identical dipole structures and switch states.}
	\label{doublesim}
\end{figure}

To verify the theoretical predictions about the transmission efficiency limit of cascaded symmetric RTA elements, a simple double-layer dipole structure is modeled and simulated in Ansys HFSS 2020. The dimensions of the structure are depicted in Fig. \ref{doublesim}, and the period is 30 mm. The simulation frequency is 9.2 GHz, and the switch is modeled as variable capacitance ranging from 1 fF to 1000 fF.

\begin{figure}[!t]
	\centering
	\subfigure[] {
		\includegraphics[width=0.46\columnwidth]{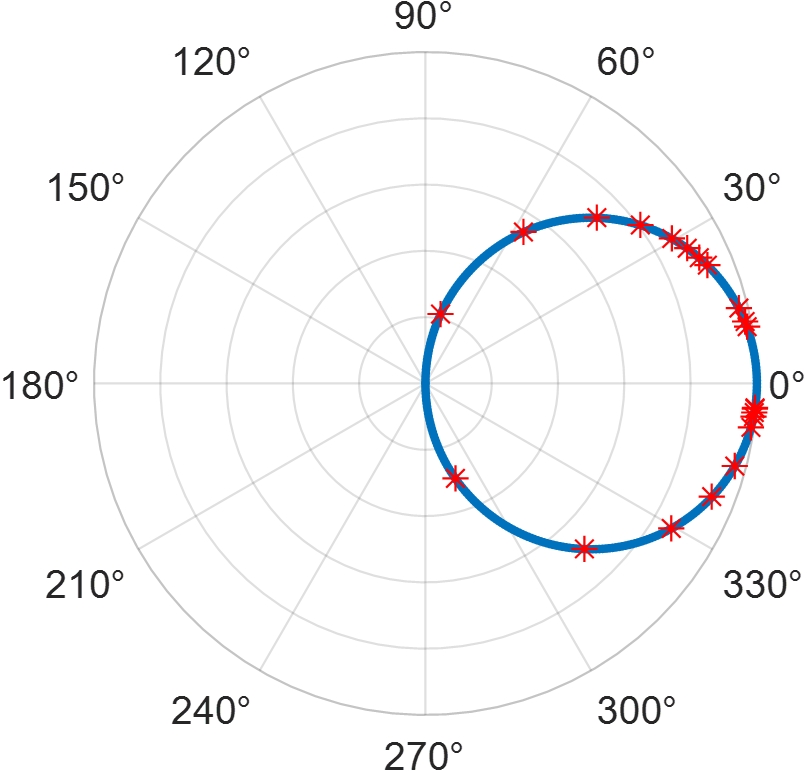} 
	} 
	\subfigure[] {
		\includegraphics[width=0.46\columnwidth]{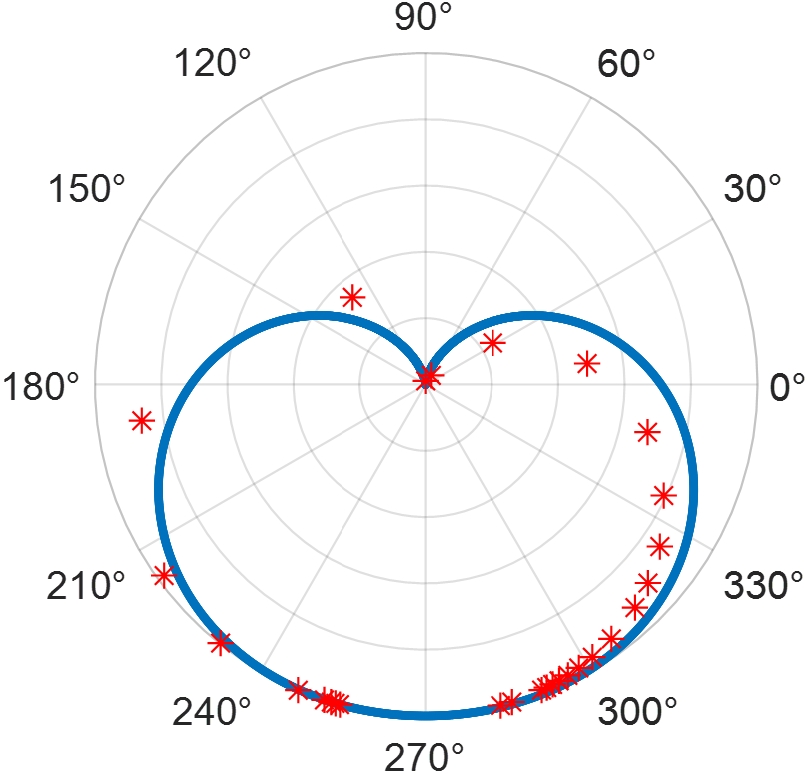} 
	} 
	\caption{Simulation results of the symmetric RTA elements. (a) Varying range of transmission coefficients for the single-layer dipole structure with a single switch. (b) Varying range of transmission coefficients for the cascaded double-layer structure. Red star: simulated transmission coefficients; Blue curve: analytical results.}
	\label{doubleresult}
\end{figure}

Initially, a single-layer dipole structure is simulated, with results shown in Fig. \ref{doubleresult}(a). Consistent with previous analysis for single-switch RTA elements, the transmission coefficients trace a unit circle with $\varphi$ = 0. Subsequently, by cascading the two identical dipole structures, a double-layer structure is formed, with each layer equipped with a switch. For simplicity, the two layers are spaced by an air gap of 8.8 mm, equivalent to $\beta D$ = 0.54$\pi$ at 9.2 GHz. The simulation results, depicted in Fig. \ref{doubleresult}(b), closely align with theoretical predictions, although discrepancies arise mainly due to mutual coupling between the layers and the influence of higher order modes. Notably, some switch states achieve a 180$^\circ$ phase difference with large transmission amplitudes, confirming the applicability of the theoretical model and the potential of cascaded configurations to enhance RTA performance.

\subsection{Analysis for Double-Layer Symmetric RTA Element with Different Symmetric Structures}

\begin{figure}[!t]
	\centerline{\includegraphics[width=0.8\columnwidth]{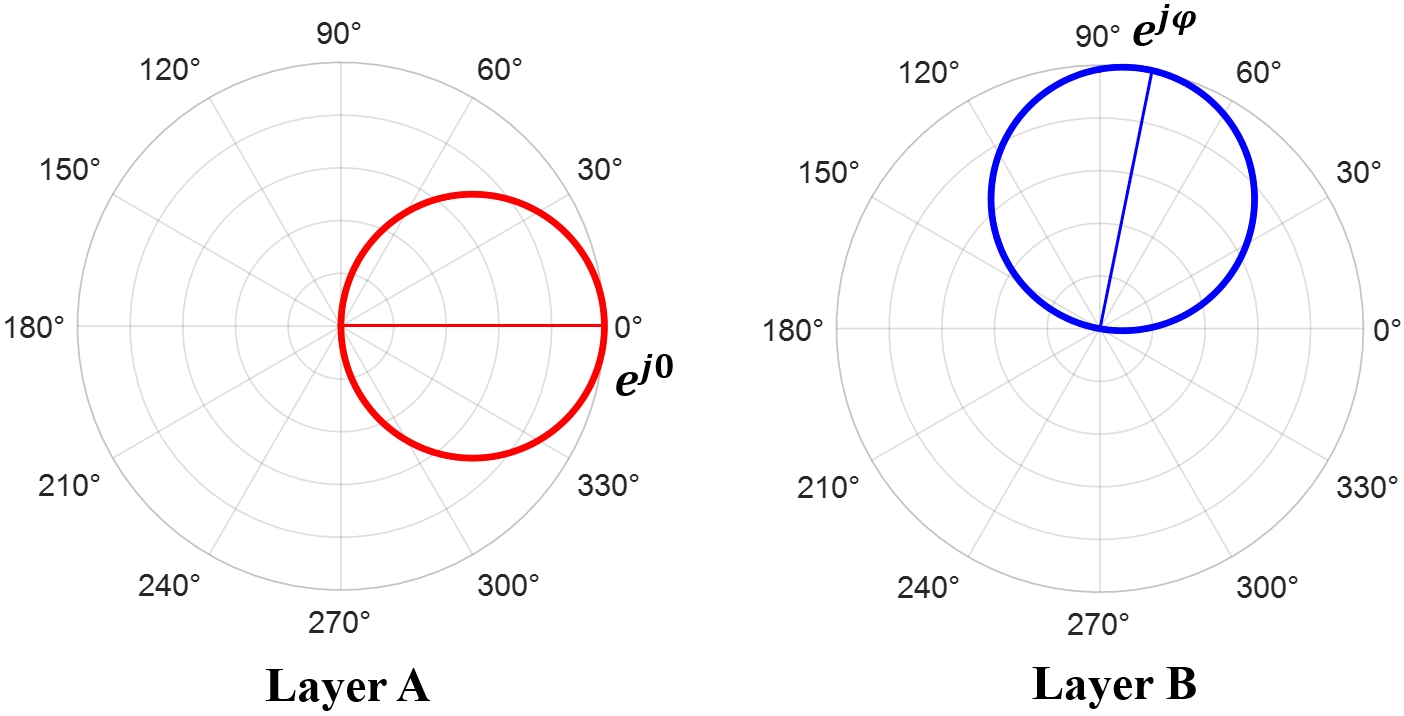}}
	\caption{The varying range for transmission coefficients of two cascaded symmetric layers with different structures and states.}
	\label{layerabdiff}
\end{figure}

Extending the previous discussions where layers A and B have identical structures and synchronous switch changes, this part explores a general scenario where the two layers possess different symmetric structures. This variation means that on the Smith chart, the transmission coefficients for the two layers trace different circles, as depicted in Fig. \ref{layerabdiff}. Without losing generality, it is assumed that the transmission coefficients of layer A vary on a unit circle with  $\varphi$ = 0, while those of layer B vary on a unit circle defined by an independent variable $\varphi$.

\begin{figure}[!t]
	\centering
	\subfigure[] {
		\includegraphics[width=0.45\columnwidth]{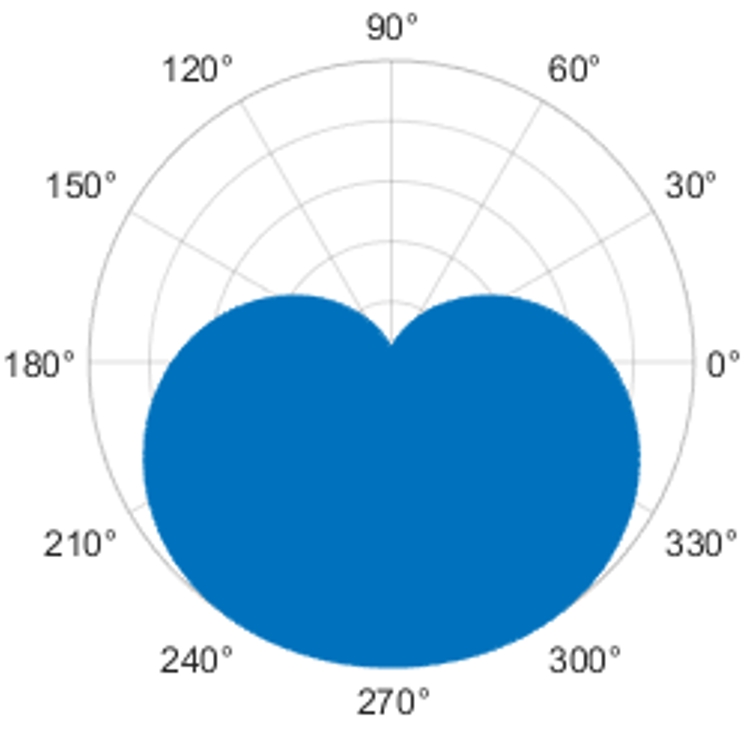} 
	} 
	\subfigure[] {
		\includegraphics[width=0.45\columnwidth]{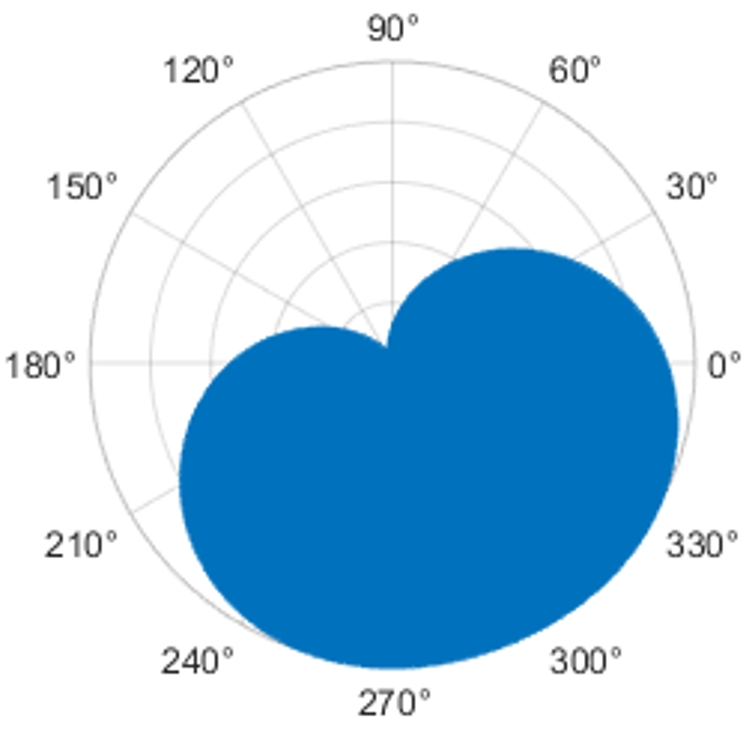} 
	} 
	\subfigure[] {
		\includegraphics[width=0.45\columnwidth]{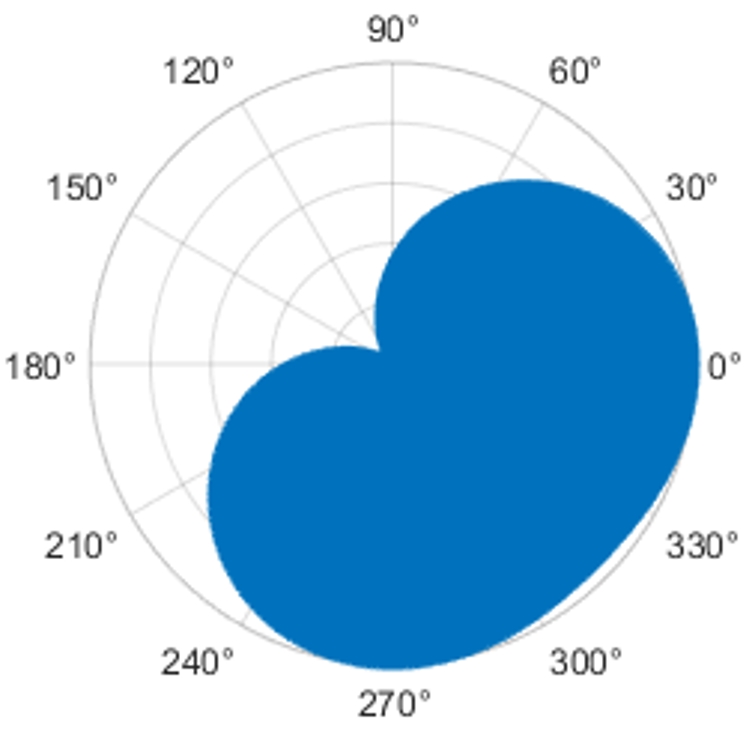} 
	} 
	\subfigure[] {
		\includegraphics[width=0.45\columnwidth]{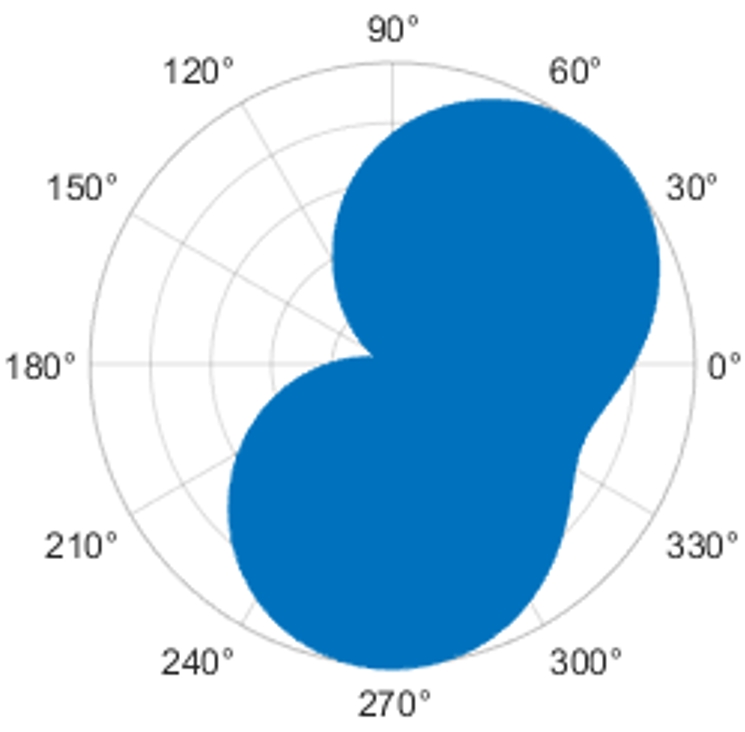} 
	} 
		\caption{Potential varying regions of the transmission coefficients with different symmetric structures in a cascaded double-layer RTA element. (a) $\varphi=0$. (b) $\varphi=\frac{\pi}{4}$. (c) $\varphi=\frac{\pi}{2}$. (d) $\varphi=\frac{3\pi}{4}$.}
	\label{symdiff}
\end{figure}

For a given value of $\varphi$, all possible transmission coefficients for the cascaded layers are numerically enumerated. The substrate thickness remains fixed at $\beta D$ = $\pi$/2 and the permittivity at $\epsilon$ = 1. The potential variations in the transmission coefficients are illustrated in Fig. \ref{symdiff}. As $\varphi$ increases, the region containing the transmission coefficients rotates and flattens. Nevertheless, across all $\varphi$ values, the transmission coefficients are confined within certain heart-shaped curves similar to those shown in Fig. \ref{dualsym}. Hence, even with different symmetric structures and varying switch states, the region where the transmission coefficients can vary remains within specific heart-shaped areas. Consequently, the performance of different symmetric layers does not exceed that of a double-layer element with identical symmetric structures and synchronous switch states.

\subsection{Inferences for Two Asymmetric Structures and Multi-Layer Structures}

From the discussions above, several conclusions can be drawn for more general configurations. Firstly, as demonstrated in Section III, for a single-layer structure with a single switch, the transmission coefficients of an asymmetric configuration are confined within the circle defined by the corresponding symmetric structures. Consequently, when cascading two asymmetric structures, it can be inferred that the varying range of transmission coefficients will also be constrained within the heart-shaped curves typical of symmetric double-layer structures. This implies that the transmission efficiency limit is established by symmetric double-layer configurations, and cannot be exceeded with asymmetric structures. 

In summary, among general cascaded double-layer structures, the configurations with identical symmetric structures and synchronous switch states have the best performance, which can guide the design of the high-performance cascaded RTA elements.

Furthermore, extending these discussions to triple-layer and multi-layer configurations, the behavior of transmission coefficients can be predicted using similar methodologies by applying the cascading S matrix formula, as seen in \cite{FSS}. Hence, with multi-layer structures, the phase variation range can be further expanded while maintaining high transmission amplitudes. This capability is advantageous for designing RTA elements that can generate continuous phase states.

\section{Conclusion}

This paper presents a theory on the transmission efficiency limit of single-switch RTA elements and their cascaded extensions. Utilizing microwave network analysis, the study employs analytical derivations and numerical simulations to demonstrate that the transmission coefficients under two states for a single-switch RTA element should lie on or inside a unit circle with a diameter of 1 on the Smith chart. This result provides solid evidence that it is not feasible to achieve both high transmission amplitudes and large phase tuning range using a single-switch RTA element. However, cascading two single-switch layers allows the transmission coefficients to vary within or on heart-shaped curves. Therefore, this configuration offers the potential to achieve high transmission amplitudes with large phase differences, which can guide the design of high-efficiency 1-bit RTA elements. The transmission efficiency limits for various scenarios discussed in this paper are summarized in Table \ref{tab}.

\begin{table}[!t]
	\centering
	\caption{The varying range of transmission coefficients under different conditions}
	\renewcommand\arraystretch{1.5}
	\scalebox{0.9}{
	\begin{tabular}{|c|l|l|}
		\hline
		\multicolumn{2}{|c|}{\textbf{Condition}} & \multicolumn{1}{|c|}{\textbf{Limit}} \\ \hline
		\multirow{2}{*}{Single-layer} & Symmetric & On a unit circle \\\cline{2-3}
		& Asymmetric & Inside a unit circle \\ \hline
		\multirow{3}{*}{Double-layer} & Symmetric, same structures  & On a heart-shaped curve \\ \cline{2-3}
		&Symmetric, different structures  & Inside a heart-shaped curve \\\cline{2-3}
		&Asymmetric & Inside a heart-shaped curve \\\hline
	\end{tabular}
}
	\label{tab}
\end{table}

Moreover, these findings are not limited to single-switch RTA elements, but also apply to other reconfigurable elements with the same three-port network topology \cite{rtatheory}. Examples of such elements include single-switch RRA elements with polarization conversion, the single-switch RTA elements based on near-field feeding method, planar reconfigurable elements with a single switch based on transmission-line feeding method, and so on.

\end{document}